\documentclass[11pt,a4paper]{article}
 \pdfoutput=1
\usepackage{jcappub}
\usepackage{amsmath,amssymb}
\usepackage{physics}
\usepackage{graphicx}
\usepackage{dcolumn}
\usepackage{bm,color}
\usepackage{hyperref}
\usepackage{accents}
\usepackage{amssymb,float}
\usepackage{amsmath}
\usepackage{multirow}
\usepackage{siunitx}
\usepackage{tabularx}
\usepackage{subcaption}
\usepackage{booktabs}
\usepackage{url}
\usepackage[shortlabels]{enumitem}

\DeclareSIUnit\parsec{pc} 
\hypersetup{
    colorlinks=true,
    linkcolor=red,
    filecolor=magenta,      
    citecolor=blue
}

\newcommand{\omp}[1]{\Omega_{#1}}

\begin{document}

\title{Constraining dark energy cosmologies with spatial curvature using Supernovae JWST forecasting}

\author[a]{Pablo M. Maldonado Alonso,}
\author[a]{Celia Escamilla-Rivera,} 
\author[a]{and Rodrigo Sandoval-Orozco}

\affiliation[a]{Instituto de Ciencias Nucleares, Universidad Nacional Aut\'{o}noma de M\'{e}xico, Circuito Exterior C.U., A.P. 70-543, M\'exico D.F. 04510, M\'{e}xico}

\emailAdd{pablo.maldonado@correo.nucleares.unam.mx}
\emailAdd{celia.escamilla@nucleares.unam.mx} 
\emailAdd{rodrigo.sandoval@correo.nucleares.unam.mx}

\date{\today}

\abstract{
Recent cosmological tensions, in particular, to infer the local value of the Hubble constant $H_0$, have developed new independent techniques to constrain cosmological parameters in several cosmologies. Moreover, even when the concordance Cosmological Constant Cold Dark Matter ($\Lambda$CDM) model has been well constrained with local observables, its physics has shown deviations from a flat background. 
Therefore, to explore a possible deviation from a flat  $\Lambda$CDM model that could explain the $H_0$ value in tension with other techniques, in this paper we study new cosmological constraints in spatial curvature dark energy models. Additionally, to standard current Supernovae Type Ia (SNIa) catalogs, we extend the empirical distance ladder method through an SNIa sample using the capabilities of the James Webb Space Telescope (JWST) to forecast SNIa up to $z \sim 6$, with information on the star formation rates at high redshift.
Furthermore, we found that our constraints provide an improvement in the statistics associated with $\omp{m}$ when combining SNIa Pantheon and SNIa Pantheon+ catalogs with JW forecasting data.
}

\maketitle

\section{\label{sec:intro}Introduction}

The first direct evidence of the late time cosmic acceleration was obtained through measurements of Type Ia Supernovae (SNIa) \citep{SupernovaCosmologyProject:1998vns, SupernovaSearchTeam:1998fmf}. Over the years, subsequent observations confirmed this result, such as the cosmic microwave background (CMB) \cite{Boomerang:2000jdg}, baryon acoustic oscillations (BAO) \cite{SDSS:2005xqv, BOSS:2013rlg}, and weak gravitational lensing \cite{DES:2017myr}. However, the capability of supernovae to prove the accelerating expansion remains invaluable since these objects are bright enough to be seen at large distances. Furthermore, SNIa are common enough to be found in large quantities, and their properties make them standardized with a precision of $\sim 0.1$ mag in brightness or $\sim 5 \%$ in distance per object \cite{Brout:2022vxf}. Also, the increasing number of SNIa observations has considerably reduced the associated statistical errors and the uncertainties in estimating cosmological parameters dominated by them \cite{Scolnic:2019apa, Lu:2022utg}. 

Nevertheless, the nature of this cosmic acceleration is one of the current inquiries in precision cosmology since still we do not fully understand the component with which it is associated, the \emph{dark energy}. However, due to the well-constrained $\Lambda$-Cold Dark Matter ($\Lambda$CDM) model, this dark energy could be evidence for a component with a negative Equation--of--State (EoS) constant value \cite{Escamilla-Rivera:2019aol, Escamilla-Rivera:2016qwv,Escamilla:2023oce} or a dynamical EoS \cite{Escamilla-Rivera:2021boq, Escamilla-Rivera:2019hqt,Zhang:2023zbj}.
Furthermore, dark energy can be associated with components that can be derived from first principles in alternative theories of gravity \cite{Jaime:2018ftn, Clifton:2011jh,Amendola:2006we}
and extended theories of gravity \cite{Bahamonde:2021gfp,Cai:2015emx}, showing a late cosmic acceleration.

On the nature of dark energy, several missions have been working to find better cosmological constraints adjoint with better systematics and increasing data baselines. Some of them as large-scale structure (LSS) observations with measurements from the Dark Energy Survey (DES) \cite{DES:2016jjg}, the Dark Energy Spectroscopic Instrument (DESI) \cite{DESI:2016fyo}, the Legacy Survey of Space and Time (LSST) on the Vera Rubin Observatory \cite{LSSTDarkEnergyScience:2018jkl}, and Euclid \cite{Amendola:2016saw}, among others, have extended the concordance cosmological model to include EoS parameters of dark energy with some shifts within 1$\sigma$.

In particular, the recently launched James Webb Space Telescope (JWST) is a very interesting experiment that can help to elucidate the nature of dark energy. JWST is a 6.5-meter primary mirror, space-based observatory operating in the visible and infrared light equipped with four main science instruments including a near-infrared camera, spectrograph, and imager slitless spectrograph; and a mid-infrared camera and spectrograph \cite{Gardner:2006ky}. It is expected that it will have an estimated lifespan of 20 years in which the research will be focused on several astrophysical and cosmology areas such as galaxy formation in the early universe as \cite{asada2022jwst,yoon2022alma,Robertson:2022gdk,harikane2023comprehensive}, exoplanet detection  \cite{luhman2023jwst,bean2023high,greene2023thermal}, metallicity and chemical exploration \cite{lai2022goals,schwieterman2022evaluating,mukherjee2022probing}, and life detection \cite{leung2022alternative,mikal2022detecting}. All these potential and current observations could allow us to explore physics further than before as testing dark energy models with structure formation \cite{Wang:2023ros,Adil:2023ara}, corroborate the Cepheid calibrations in the distance ladder \cite{Riess:2023bfx} and adding more SNIa observations \cite{Lu:2022utg}, cosmic chronometers \cite{Cogato:2023atm}, and XA/UV quasars \cite{Dainotti:2023cpn,Sandoval-Orozco:2023pit} to the constraint analysis of them. 

Recently, numerous studies related to the implications of JWST to cosmology have been developed. In \cite{Lu:2022utg} a JWST simulated sample of SNIa within a redshift range  $2 \lesssim z \lesssim 6$, was employed to constrain standard cosmological parameters. Using combinations of the mock sample and SN Pantheon dataset \cite{Pan-STARRS1:2017jku} it was possible to constrain dark energy models with constant EoS. This analysis was performed using two different forms of the intrinsic evolution of the standardized SNIa luminosity distance. On one hand, it is assumed a linear redshift dependence of magnitude evolution. On the other hand, we can consider other logarithmic evolutions. Analysing the cases with and without systematic evolution, it was found that the addition of the simulated SNIa sample would successfully remove the evolutionary effects.  However, even though the SN Pantheon dataset size was increased by a factor of $\approx$ 16 data points, it is still not able to constrain the systematic evolution and the cosmological parameters as effectively as the very high redshift SN data.

According to several works \cite{2023Atek, 2023ApJ...948L..13J, 2023ApJ...942L...9Yan, 2022ApJ...938L..15Castellano, 2022A&A...665L...4Schaerer,2022MNRAS.511.5475Marshall, 2022ApJ...937L..33Suess}, there are some common aspects 
within the structure morphology, which indicate that galaxies discovered by JWST do not have enough time to evolve into what it is observed today. The first JWST measurements at very high redshift aim at the existence of a population of massive galaxies with stellar masses $\geq 10^{10.5}$ $M_{\odot}$ \cite{Finkelstein_2023, Treu_2022, Santini_2023, Castellano_2022}. This result would indicate a higher cumulative stellar mass density than predicted by the $\Lambda$CDM model in the redshift range $7<z<11$ \cite{labbe2022very}. Furthermore, a quantification of the tension between the standard cosmological theory and JWST observations was recently carried out in \cite{Wang:2023gla}. However, in comparison to previous works, the estimations computed so far are based on the statistics of the total numbers of massive halos and not on the statistics of the cumulative stellar mass. Therefore, the results reported in \cite{Wang:2023gla} show a 2-3$\sigma$ tension with the $\Lambda$CDM cosmology for a low star formation efficiency $\sim 0.1$. Furthermore, it is proven that for a star formation efficiency at higher redshift $z \gtrsim 0.3$, the JWST galaxy measurements are consistent with the $\Lambda$CDM model  \cite{Labb_2023}. This approach can be considered in our mock analysis in the redshift range indicated since the possibility of using a standard prior cosmology on the simulation can be reasonable at this scale. Moreover, it is important to mention that any model assumption on the simulations outside this redshift range should take into account the possibility of a dynamical cosmological model.

A further study about the first galaxies discovered by JWST was carried out in \cite{Lovyagin:2022kxl}. It is important to notice that this study is within the framework of the standard cosmological $\Lambda$CDM model. Therefore, the JWST dataset includes near and mid-infrared images and near-infrared spectra to perform analyses based on cosmographic theories of the angular size-redshift relationship \cite{Lovyagin:2022kxl}. The $\Lambda$CDM interpretation of JWST observations is compared with the interpretation based on Zwicky's static universe model \cite{Zwicky:1929zz}, where the origin of the cosmological redshift can be explained through the photon-energy loss mechanism. However, the redshifted objects detected by the JWST are not aligned with such interpretation, but before any conclusion, data from this mission should be increased.

In \cite{Forconi:2023hsj} the recent tension related to the JWST observations of a population of very high-mass galaxies is faced through the correlation between JWST data and $\Lambda$CDM cosmological parameters. Under this approach, neglecting some systematic effect, it is argued that for models with a larger matter density value, a higher amplitude of primordial inflationary fluctuations and a bigger scalar spectral index can predict larger cumulative stellar mass densities and, as a result, provide a better fit to the JWST dataset. In this context, theories related to the dark energy sector, especially, early dark energy models, seem to be good candidates to explain the problem of massive galaxies observed by JWST.

Currently, the implementation of modern and powerful surveys has allowed us to obtain data from our Universe with greater precision at increasingly higher redshifts. The statistical errors associated with the new SN Ia measurements have been reduced such that, in some cases, the uncertainties of the cosmological parameters are largely due to systematic errors \cite{Riess:2021jrx}. In this regard, more precise constraints on parameters could be achieved by focusing our efforts on reducing systematic errors. However, this does not rule out that it is possible to obtain improvements in uncertainties only with the addition of high redshift SN Ia data. It has been proved that at high redshift ($z>2$) SNe Ia implementation led us to a significant improvement of the matter density and vacuum density errors for $\Lambda$CDM cosmology by being reduced $73\%$ and $67\%$, respectively \cite{Lu:2022utg}. Also, in the Chevallier-Polarski-Linder model, an error decrease for the matter density ($70 \%$), $w_0$ ($29 \%$) and $w_a$ ($55 \%$) is reported. Similarly, in \cite{Salzano:2013jia} the addition of $\sim 36$ JWST SN Ia at $1.5<z<3.5$, reduce the errors on Chevallier-Polarski-Linder model parameters ($\sigma_{w_0}$ and $\sigma_{w_a}$) approximately $\sim 9 \%$ and $\sim 14 \%$, respectively. Thus, for models like the previous one, SN Ia distance measurements are more sensitive to certain components of the EoS than others \cite{Salzano:2013jia}. Therefore, it is interesting to explore the impact that the high redshift SN Ia attainable by JWST has on each of the dark energy EoS, as well as to measure the degree of influence on the constriction of cosmological parameters at hand.

As a step forward, using the capabilities of the JWST described, in this work, we develop the forecasting of SNIa up to $z \sim 6$, with information on the Star Formation Rates (SFR) at high redshift. Once this data is at hand, we perform a statistical analysis combined with SN Pantheon \cite{Pan-STARRS1:2017jku} and SN Pantheon+\footnote{\href{https://github.com/PantheonPlusSH0ES/DataRelease}{github.com/PantheonPlusSH0ES/DataRelease}} to constrain spatial curvature dark energy cosmologies. We based our cosmological models inspired by bidimensional EoS parameterisation, which preserves the expanding and accelerating behaviour at late times. Our goal is to show that a simple deviation in the spatial curvature of a dark energy EoS model can verify a well-constrained analysis with SNIa JWST forecasting.

This paper is divided as follows:
In Sec.~\ref{sec:darkenergy} we summarise the theory behind dark energy bidimensional parameterisations inspired in the Taylor series around the cosmological scale factor $a$. All of these parameterisations are described through their normalised $E(z)$ Friedmann evolution equation, including the curvature term. Furthermore, we are going to consider standard $\Lambda$CDM and $w$CDM models in addition to the dark energy cosmologies to proceed with comparisons between them. Also, we include the latest constraints reported in the literature so far.
In Sec.~\ref{sec:data} we present the methodology employed for observables. We include the description of current SNIa data baselines and how we can proceed with their forecasting using JWST characteristics. In Appendix \ref{sec:Appendix-A} we describe the technicalities behind this forecasting.
The results on new constraints for the models described are developed in Sec.~\ref{sec:Results}.
Finally, the conclusions are presented in Sec.~\ref{sec:conc}.

\section{Standard dark energy parameterisations}
\label{sec:darkenergy}

The standard cosmological scenario $\Lambda$CDM is a remarkable fit for most cosmological data. Nonetheless, we are still searching for the nature of inflation, dark matter, and dark energy. Physical evidence for these components comes only from astrophysical and cosmological observations \cite{DiValentino:2021izs}. Therefore, an increase in experimental sensitivity can produce deviations from the standard $\Lambda$CDM scenario that could lead to a deeper understanding of the gravity theory. If it is not a consequence associated with systematic errors, the cosmological tensions \cite{Abdalla:2022yfr}
existing between the different experimental probes could indicate a failure of the $\Lambda$CDM model, and a better cosmological model should be able to be found. 

In this section, we are going to describe, in addition to the $\Lambda$CDM model with EoS $w=-1$, five dark energy bidimensional in $z$ parameterisations of $w(z)$, which can be constant or redshift-dependent. Notice that to describe dark energy, we need to achieve cosmic acceleration with a negative pressure at late times \cite{Escamilla-Rivera:2016qwv}. Furthermore, in \cite{Escamilla-Rivera:2016aca} was discussed parameter constraints from the full-shape CMB power spectrum for each parameterisation, however, we describe in the following lines an update for each of them. Also, in \cite{Zhao:2020ole} it was analysed $\Lambda$CDM and CPL models using Fast radio bursts combined with the CMB shape spectra showing an improvement in dark energy parameter constraints. Other datasets have been used in testing the aforementioned models as the \textit{galaxy full-shape} data and the Baryon Acoustic Oscillations in both \citep{DAmico:2020kxu,Chudaykin:2020ghx}.


\begin{itemize}

\item \textbf{$\Lambda$CDM model.}
In this model, the universe is composed of cosmological fluids with different EoS's $w$ that contribute to the energy constraint. At present cosmic times, the non-relativistic matter contribution, $\omp{m} \simeq 0.27$, is the sum of the ordinary baryonic matter term, $\Omega_b \simeq 0.044$, and the (cold) dark matter term, $\Omega_{c} \simeq 0.22$. Dark energy ($\Omega_{\Lambda} \simeq 0.73$) is described by $w=-1$, associated with a cosmological constant $\Lambda$ or a vacuum energy density \cite{amendola_tsujikawa_2010}. Radiation represents a negligible contribution, $\Omega_r \simeq 9\times 10^{-5}$, but it dominated the early cosmic stages, after the end of the inflationary stage and before matter-radiation decoupling \cite{soton420204}.
Additionally, $\Lambda$CDM can be characterised with a flat geometry, which corresponds to an energy density parameter, $ \omp{\Lambda} = 1-\omp{m}-\omp{r}$, where the only parameter to be constrained is $\omp{m}$. The cosmological evolution for this model can be expressed as
\begin{equation}
        E (z) \equiv \dfrac{H(z)}{H_0} =
\sqrt{ \omp{m} (1+z)^3 + \omp{r} (1+z)^4 + \omp{\Lambda} },
\label{E(z)2}
\end{equation}
where $H_0$ is the Hubble constant today.

We also consider the non-flat $\Lambda$CDM cosmological model, as an extension of the $\Lambda$CDM model but with curvature $k \neq 0$, with its constraint equation as $\omp{k} = 1- \omp{m} - \omp{r} - \omp{\Lambda}$, where  $\Theta = (\omp{m}, \omp{\Lambda})$ is the vector with free parameters. The evolution for this case can be written as
\begin{equation}
        E (z) =
\sqrt{ \omp{m} (1+z)^3 + \omp{r} (1+z)^4 + \omp{k} (1+z)^2 + \omp{\Lambda} }.
\label{E(z)}
\end{equation}
Then, the model-dependent luminous distance can be calculated according to the $\omp{k}$ value \cite{Bargiacchi:2021hdp}:

\begin{equation}
    D_L (z) = \left\lbrace
\begin{array}{ll}
\dfrac{c}{H_0} (1+z) \dfrac{\sinh{ \left( \sqrt{\omp{k}} \int_{0}^{z}\dfrac{dz'}{E(z')} \right) }}{\sqrt{\omp{k}}}, &\quad\quad \omp{k} > 0 \\

\dfrac{c}{H_0} (1+z) \int_{0}^{z} \dfrac{dz'}{E(z')},  &\quad\quad  \omp{k} = 0 \\

\dfrac{c}{H_0} (1+z) \dfrac{\sin{ \left( \sqrt{-\omp{k}} \int_{0}^{z} \dfrac{dz'}{E(z')}  \right) }}{\sqrt{- \omp{k}}}, &\quad\quad  \omp{k} < 0,
\end{array}
\right.
\label{dl(z)}
\end{equation}
where $c$ the speed of light and $E(z)$ the background evolution equation of the cosmological models.

The base test for $\Lambda$CDM is the analysis provided by the Planck collaboration measuring CMB anisotropies finding the base parameters values $\Omega_m= 0.316 \pm 0.008$ and $H_0 = 67.27 \pm 0.6$ km s$^{-1}$ Mpc$^{-1}$ \citep{Planck:2018vyg} in the context of a flat $\Lambda$CDM. Using late-time data in \cite{Bargiacchi:2021hdp} was found that using SNIa, BAO, and a quasar sample the values are $\Omega_m = 0.300 \pm 0.012$, with a fixed $H_0 = 70$ km s$^{-1}$ Mpc$^{-1}$, while using a non-flat $\Lambda$CDM model the results were $\Omega_m = 0.364 \pm 0.021$ and $\Omega_\Lambda = 0.829 \pm 0.035$, finding a slight deviation from the flat background. Furthermore, with a Cosmic Chronometers (CC -- $H(z)$) sample it was found that $H_0 = 66.7 \pm 5.3$ km s$^{-1}$ Mpc$^{-1}$ and $\Omega_m = 0.33^{+0.08}_{-0.06}$ for the same flat model \cite{Moresco:2023zys}. Using only SNIa Pantheon+ compilation \citep{Brout:2022vxf} the values for the flat $\Lambda$CDM model are $H_0 = 73.6 \pm 1.1$ km s$^{-1}$ Mpc$^{-1}$. The latter assumes a Gaussian prior with $\Omega_m = 0.334 \pm 0.018$. While for the non-flat $\Lambda$CDM the results are $\Omega_m = 0.306 \pm 0.057$ and $\Omega_\Lambda = 0.625 \pm 0.084$ in concordance with the flat counterpart at $2\sigma$. Using the Planck collaboration measurements for a non-flat $\Lambda$CDM model results in a $\Omega_k = -0.044^{+0.018}_{-0.015}$ which is an apparent curvature result over $2\sigma$. Joining the CMB spectrum and the BAO results obtains $\Omega_k = 0.0007 \pm 0.0019$ \citep{Planck:2018vyg}. Employing galaxy full-shape data (FS), BAO and SNIa results in $\Omega_k = -0.043^{+0.036}_{-0.036}$ in \citep{DAmico:2020kxu}.


\item \textbf{$w$CDM model.}
The simplest extension of the $\Lambda$CDM model is the one in which $w \neq -1$, yet still constant in time meaning that $w(z) = w_0$. From \cite{Lu:2022utg,Bargiacchi:2021hdp} we express $E(z)$ for this model as,
\begin{equation}
    E(z) = \sqrt{ \omp{m} (1+z)^3 + \omp{k} (1+z)^2 +\omp{\Lambda} (1+z)^{3(1+w_0)} }.    
\end{equation}
Non-flat $w$CDM has $\Theta = (\omp{m}, \omp{\Lambda}, w_0)$ as free parameters. However, under flatness assumption, $\omp{k}$ = 0, so the only free parameters are $\Theta = (\omp{m}, w_0)$. This model reduces to $\Lambda$CDM when $w_0 = -1$.

Using SNIa, BAO, and quasars the values obtained in \cite{Bargiacchi:2021hdp} were: $\Omega_m = 0.369^{+0.022}_{-0.023}$ and $w_0 = -1.283^{+0.094}_{-0.027}$, corresponding to a deviation from the $\Lambda$CDM model in more than $1\sigma$ range with the same fixed $H_0$ value. While using a non-flat $w$CDM results in $\Omega_m=0.280^{+0.041}_{-0.037}$, $\Omega_\Lambda = 1.662^{+0.041}_{-0.048}$ and $w_0 = -0.667^{+0.024}_{-0.027}$, where a difference from a flat model is reported of more than $3\sigma$ using only SNIa and quasars. Furthermore, adding BAO to the quasar sample \citep{Zheng:2021oeq} results in $\Omega_m = 0.31 \pm 0.03$ and $w_0 = -1.00^{+0.14}_{-0.13}$, that is consistent with $\Lambda$CDM assuming a Gaussian prior of $H_0 = 67.32 \pm 4.7$ km s$^{-1}$ Mpc$^{-1}$. When using only SNIa \citep{Brout:2022vxf} the flat $w$CDM model gives $\Omega_m = 0.309^{+0.063}_{-0.069}$, $H_0 = 73.5 \pm 1.1$ km s$^{-1}$ Mpc$^{-1}$ and $w_0 = -0.90 \pm 0.14$, returning the confirmation for a $\Lambda$CDM model. The analysis done using the Planck measurements obtains $w_0 = -1.028 \pm 0.031$ for a flat model, in a $2\sigma$ range from the $\Lambda$CDM model \citep{Planck:2018vyg}. A comparison between the CMB power spectrum and the growth of structure data like BAO, $f\sigma_8$ and lensing is done in \citep{Andrade:2021njl} for the $w$CDM model that results in $w_0 = -1.004^{+0.096}_{-0.085}$ and $w_0 = -1.043^{+0.087}_{-0.069}$. Using the FS data, BAO and SNIa samples results in $w_0 = - 1.031^{+0.052}_{-0.048}$ \citep{DAmico:2020kxu}, While employing the large scale structure, the results presented in \citep{Chudaykin:2020ghx} obtain $w_0 = -1.046^{+0.055}_{-0.052}$.


\item \textbf{Chevallier–Polarski–Linder (CPL) model.}
One of the most used redshift-dependent parameterisations corresponds to the Chevallier-Polarski-Linder \cite{Chevallier:2000qy, Linder:2002et} proposal: $w(z) = w_0 + w_a z/(1+z)$. In which, $w(z) = w_0 + w_a$ at $z=\infty$ and $w(z) = w_0$ at $z=0$, but it diverges in the future for $z \rightarrow (-1)^{+}$. In this bidimensional model, $w_0$ denotes the dark energy EoS today, and $w_a$ describes its evolution. This parameterisation has several advantages including the well behaviour at high redshift, the linear feature at low redshift, a simple physical interpretation, and the accuracy in reconstructing a scalar field EoS \cite{Linder:2002et}. The normalised Hubble parameter for this model can be written as
\begin{equation}
    E(z) = \sqrt{ \omp{m} (1+z)^3 + \omp{k} (1+z)^2 + \omp{\Lambda} (1+z)^{3(1+w_0+w_a)}
 \exp(\dfrac{-3 w_a z}{1+z}) }.
\end{equation}
For the non-flat and flat cases, we consider as free parameters ($\omp{m}$, $\omp{\Lambda}$, $w_0$, $w_a$) and ($\omp{m}$, $w_0$, $w_a$), respectively. This model can be reduced to $\Lambda$CDM with $w_0 = -1$ and $w_a = 0$. 

Using SNIa, BAO, and quasars in \citep{Bargiacchi:2021hdp} was discussed a deviation from the $\Lambda$CDM with $\Omega_m = 0.354^{+0.032}_{-0.030}$, along with $w_0 = -1.323^{+0.103}_{0.112}$ and $w_a = 0.745^{+0.483}_{-0.974}$ for a flat CPL model, values that correspond to a confirmation of the $\Lambda$CDM model. Using quasars and $H(z)$ measurements for a non-flat CPL parametrization it was found that $\Omega_m = 0.44 \pm 0.10$, $\Omega_k = -0.36 \pm 0.24$, $H_0 = 71.8^{+4.6}_{-7.7}$, with $w_0 = -1.2 \pm 1.0$, and $w_a = -5.0^{+9-0}_{-2.0}$ \cite{Dinda:2023svr}  showing a clear deviation of more than $2\sigma$ from the flat $\Lambda$CDM model. Using SNIa \citep{Brout:2022vxf} for the flat CPL with a Gaussian prior in $H_0$ results in $H_0 = 73.3 \pm 1.1$ km s$^{-1}$ Mpc$^{-1}$, with $\Omega_m = 0.403^{+0.054}_{-0.098}$, $w_0 = -0.93 \pm 0.15$ and $w_a = -0.1^{+0.9}_{-2.0}$. This latter corresponds to a flat $\Lambda$CDM confirmation. Furthermore, adding BAO, CMB to the previous SNIa sample results in $H_0 = 67.41^{+0.52}_{-0.82}$ km s$^{-1}$ Mpc$^{-1}$, $\Omega_m = 0.316^{+0.009}_{-0.005}$ and $w_0 = 1.267^{+0.196}_{-0.191}$ and $w_a = -3.771^{+2.113}_{-2.496}$ \cite{Bargiacchi:2021hdp}. In \citep{Buenobelloso:2011sja} analyses are done using the CMB distance priors, BAO datasets and SNIa resulting in $w_0 = -0.93^{+0.073}_{-0.085}$ and $w_a = -0.63^{+0.035}_{-0.082}$ at $1\sigma$ deviation from $\Lambda$CDM. The combined results of FS, BAO and SNIa obtain $w_0 = -0.98^{+0.10}_{-0.11}$ and $w_a = -0.32^{+0.63}_{-0.48}$ presented in \citep{DAmico:2020kxu}.


\item \textbf{Jassal-Bagla-Padmanabhan (JBP) model.}
In \cite{Jassal:2004ej} the parameterisation $w(z) = w_0 + w_a z/(1+z)^2$, in which $w(0) = w_0$, $w'(0) = w_1$, and $w(\infty) = w_0$ is presented based on trying to explain the accelerated universe covering both CMB and the SNIa measurements. This model was proposed to solve the high $z$ issues within the CPL parameterisation \citep{Escamilla-Rivera:2016qwv}. Using the behaviour of this function allows us to have the same EoS at the present epoch and high-$z$ with a rapid variation at small redshifts. Considering the corresponding term related to curvature, the derivative expression for $E(z)$ for this model can be written as
\begin{equation}
    E(z) = \sqrt{ \omp{m} (1+z)^3 + \omp{k} (1+z)^2 + \omp{\Lambda} (1+z)^{3(1+w_0)} 
 \exp[ \dfrac{3 w_a z^2}{2(1+z)^2} ] },
\end{equation}
where $\Theta = (\omp{m}, \omp{\Lambda}, w_0, w_a)$ for the non-flat JBP model, and $\Theta = (\omp{m}, w_0, w_a)$ for a flat JBP model.

Using SNIa, BAO, and quasars the JBP model is tested obtaining $\Omega_m = 0.354^{+0.032}_{-0.030}$, $w_0 = -1.371\pm 0.141$ and $w_a = 1.127^{+1.293}_{-1.547}$ \citep{Bargiacchi:2021hdp} which is considered a deviation from the confirmation for $\Lambda$CDM for the $w_0$ value. In \citep{Staicova:2022zuh} using SNIa, BAO, CMB and Gamma-Ray Bursts (GRB) it is obtained $\Omega_m = 0.27 \pm 0.03$ with $w_0 = -1.02 \pm 0.04$ and $w_a =  0.22 \pm 0.23$ using a flat JBP model in concordance with a flat $\Lambda$CDM at $1\sigma$. In \citep{Buenobelloso:2011sja} an analysis using the CMB distance priors, BAO and SNIa measurements obtained $w_0 = -0.78^{+0.027}_{-0.107}$ and $w_a = -1.12^{+0.164}_{-0.251}$ with more than $1\sigma$ deviation from the $\Lambda$CDM.


\item \textbf{Exponential model.}
In \cite{Yang:2018qmz} was examined five one-parameter dark energy parameterisations with several datasets. In particular, data from CMB observations, Joint light-curve analysis from SNIa observations (JLA), BAO distance measurements, and $H(z)$. It was concluded that the one-parameter dark energy model can provide a solution to the $H_0$ tension between local measurements and Planck indirect ones. Besides, it was found which of the five models is better fitted to the data used. This model, relatively close to $\Lambda$CDM, is the one with an EoS of the form: $w(z) = w_0 \exp[z/(1+z)] /(1+z)$, where $w(0) = w_0$ and $w(z) = 0$, for both $z=\infty$ and $z \rightarrow (-1)^{+}$. As a result, the normalised Hubble parameter can be written as
\begin{equation}
        E(z) = \sqrt{ \omp{m} (1+z)^3 + \omp{k} (1+z)^2 + \omp{\Lambda} (1+z)^{3} 
 \exp[ 3 w_0 \left( \exp( \dfrac{z}{1+z} )-1 \right) ] }.
\end{equation}
For non-flat exponential model we have $\Theta = (\omp{m}, \omp{\Lambda}, w_0)$, while for flat exponential model, $\Theta = (\omp{m}, w_0)$.

Using SNIa, quasars, and BAO was obtained for the Exponential model $\Omega_m = 0.359^{+0.023}_{-0.024}$ and $w_0 = -1.271^{+0.092}_{-0.107}$ \citep{Bargiacchi:2021hdp}  showing again a deviation from a flat $\Lambda$CDM. In \cite{Castillo-Santos:2022yoi} the exponential model is constrained using CMB, SNIa, BAO, and measurements of the distance using Hydrogen II galaxies resulting in $H_0 = 70.9 \pm 7.0$ km s$^{-1}$ Mpc$^{-1}$, $\Omega_m = 0.284 \pm 0.006$ and $w_0 = -1.202^{+0.027}_{-0.026}$, imposing a Gaussian local prior in $H_0$. Although there is very few analysis using the exponential model with the CMB power spectrum, the model have been implemented in a modified version in \citep{Yin:2020dwl}. The extra acoustic term $c_s^2$ is used to perform the analysis. 

 
\item \textbf{Barboza-Alcaniz (BA) model.}
In \cite{Barboza:2008rh} was proposed a dark energy parameterisation given by $w(z) = w_0 + w_a z(1+z) / 1+z^2$. This is a well-behaved function of redshift throughout the entire cosmic evolution, $z \in [-1, \infty]$, with $w(z) = w_0 + w_a$ for $z = \infty$ and $w(z) = w_0$, when $z \rightarrow (-1)^{+}$. This smooth function allows to define regions on the plane $(w_0-w_1)$ associated with several classes of dark energy models to exclude or confirm the models based on the constraints using observational data. Thus, it was shown that both quintessence and phantom behaviours have fully acceptable regimes. The $E(z)$ for non-flat case of this model can be written as
\begin{equation}
        E(z) = \sqrt{ \omp{m} (1+z)^3 + \omp{k} (1+z)^2 + \omp{\Lambda} (1+z)^{3(1+w_0)} \left( 1+z^2 \right)^{\frac{3 w_a}{2}} }.
\end{equation}
The free parameter sets for the non-flat BA model and flat BA model are $\Theta = (\omp{m}, \omp{\Lambda}, w_0, w_a)$ and $\Theta = (\omp{m}
, w_0, w_a)$, respectively.
The analysis using BAO, SNIa and quasars in \cite{Bargiacchi:2021hdp} showed deviation from the flat $\Lambda$CDM as $\Omega_m = 0.307^{+0.044}_{-0.055}$, $w_0 = -1.303^{+0.115}_{-0.106}$ and $w_a = 1.010^{+0.152}_{-0.466}$, meaning that the quasar sample is responsible for a deviation from the standard model. In \cite{Staicova:2022zuh} using CMB, BAO, SNIa and GRB results in $\Omega_m = 0.28 \pm 0.03$, $w_0 = -1.13 \pm 0.04$ and $w_a = 0.37 \pm 0.1$, finding a lower deviation from the $\Lambda$CDM model, but using only BAO and CMB the standard model is recovered with $\Omega_m = 0.29 \pm 0.04$, $w_0 = -1.06 \pm 0.11$ and $w_a = 0.35 \pm 0.12$. 
In \citep{Buenobelloso:2011sja} the analysis is done using CMB distance priors, BAO and SNIa obtaining $w_0 = -0.725^{+0.082}_{-0.079}$ and $w_a = -0.43^{+0.013}_{-0.041}$ that shows again a deviation of more than $1\sigma$ from the $\Lambda$CDM model.

\end{itemize}


\section{Data treatment: observations and forecastings}
\label{sec:data}

In this section, we will perform the statistical analysis for the dark energy models with and without curvature including three different datasets: SNIa Pantheon and SNIa Pantheon+ samples, along with the extracted simulated data from JWST.

\begin{itemize}

\item \textbf{Pantheon (PN)}  \cite{Pan-STARRS1:2017jku}: 
The Pantheon compilation is a combination of measurements from SNIa distances combining both low and high redshifts from $z \sim 0.01$ up to $z = 2.26$. This sample has shown an improvement in the photometric calibrations on the distance ladder through the light curve. This transforms observable quantities into distances adding up to a total of 1048 data points. 

\item \textbf{Pantheon+ (PN$^+$)} \cite{Brout_2019,Riess:2021jrx}: 
Pantheon+ is a collection of 18 different SNIa samples based on the Pantheon compilation above described and by adding new data points recollected from different surveys as: Foundation Supernova Survey \cite{Foley:2017zdq}, the Swift Optical/Ultraviolet Supernova Archive (SOUSA) \cite{Brown:2014gqa}, the Lick Observatory Supernova Search LOSS1 \cite{Ganeshalingam_2010}, the second sample LOSS2 \cite{Stahl:2019xzs}, and DES \cite{Brout_2019}. As a result, Pantheon+ consists of 1701 light curves of 1550 distinct SNIa along a redshift of $z = $ 0.001 up to 2.26. For $H_0$, a value of $73.30 \pm 1.04$ km s$^{-1}$ Mpc$^{-1}$ is assumed. This sample is represented in Figure \ref{fig:hist1} in blue colour. 
Cosmological parameter constraints have been carried out using the affine-invariant ensemble sampler for Markov Chain Monte Carlo (MCMC) module 
\texttt{emcee}\footnote{\href{https://emcee.readthedocs.io/en/stable/}{emcee.readthedocs.io/en/stable/}} 
that uses random generation numbers to explore the parameter space based on the probability function $P \propto \exp(-\chi^{2}/2)$ to minimize the quantity

\begin{equation}\label{eq:chi2pantheon+}
    \chi^{2}_{\text{SNIa}} (\Theta) = \Delta \mu ^{T} (z, \Theta) \cdot C^{-1}_{\text{SNIa}} \cdot \Delta \mu (z, \Theta) + \ln{\left( \dfrac{S}{2 \pi} \right)},
\end{equation}
where $\Delta \mu (z, \Theta) = \mu(z)_{\text{data}} - \mu(z, \Theta)_{\text{model}}$, $C_{\text{SNIa}}$ is the covariance matrix of the PN (or PN+) sample, $S$ is the sum of all the components of $C^{-1}_{\text{SNIa}}$, $\mu(z)_{\text{data}}$ is the distance modulus of the PN (PN+) data, and $\mu(z, \Theta)_{\text{model}}$ is the distance modulus for a cosmological model with a parameter set $\Theta$ \cite{Briffa:2021nxg}.

\item \textbf{The extracted JWST simulated SN data (JW)} \cite{Lu:2022utg}:
Recently it was created an SNIa sample at $z > 2$ using a Monte Carlo simulation derived from the capabilities of JWST and based on the FLARE project \cite{Wang:2017awy, Regos:2018fgq}. The SNIa redshifts (up to $z \sim 6$) were derived by extrapolating the local supernovae rates with information on SFR at high redshift \cite{Chen:2021hjw}. With this survey, in 6 years we expect to measure 205 SNIa in a range of $z \sim 0.3$ up to $z \sim 6$. We represent this sample in Figure \ref{fig:hist1}.
The mock sample was created based on a flat-$\Lambda$CDM model using a fit from Pantheon data. To develop our analysis, we changed the sample for SN PN$^+$. 
The full covariance matrix is constructed by adding the off-diagonal value of $0.002^2$ with + and – signs assigned randomly for all the matrix. The value in the diagonal is the mean value of the Pantheon covariance matrix \cite{Scolnic:2019apa} and the $0.15^2$ value in the diagonal is the assumed error considering the local determination of the observation errors extrapolated \cite{Lu:2022utg}. The covariance matrix has the following form
\begin{equation}\label{eq:matrix_jw}
   C_{\text{JW}} = \underbrace{\begin{pmatrix}
    \ 0.15^2 & \pm0.002^2 & \pm0.002^2 & \dots & \pm0.002^2 \ \null \\
    \ \pm0.002^2 & 0.15^2 & \pm0.002^2 & \dots & \pm0.002^2 \ \null \\
    \ \pm0.002^2 & \pm0.002^2 & 0.15^2 & \dots & \pm0.002^2 \ \null \\
    \ \vdots & \vdots & \vdots & \ddots& \vdots \ \null \\
    \ \pm0.002^2 & \dots  & \dots  & \dots & 0.15^2 \ \\
    \end{pmatrix}}_{\substack{\\[0.2em] \text{\normalsize $n \times n$}}},
\end{equation}
where $n \times n$ denotes the dimension of the matrix and ($\pm$) the random assignment of (+) and (--) signs. In this case, 

\begin{equation}\label{eq:deltam}
    \chi^{2}_{\text{JW}} (\Theta) = \Delta \mu ^{T} (z, \Theta) \cdot C^{-1}_{\text{JW}} \cdot \Delta \mu (z, \Theta) + \ln{\left( \dfrac{S}{2\pi} \right)},
\end{equation}
where $\Delta \mu (z, \Theta) = \mu(z)_{\text{data}} - \mu(z, \Theta)_{\text{model}} - \Gamma_0$, $C_{\text{JW}}$ is the covariance matrix presented in Eq.(\ref{eq:matrix_jw}), $S$ is the sum of all the components of $C^{-1}_{\text{JW}}$, $\mu(z)_{\text{data}}$ is the distance modulus of the JW data, $\mu(z, \Theta)_{\text{model}}$ is the distance modulus for a cosmological model with parameter set $\Theta$ and $\Gamma_0$ is a magnitude bias added to consider a possible systematic luminosity difference between the JW and PN (PN+) datasets. Notice that if we compare our Eq.(\ref{eq:deltam}) with \cite{Lu:2022utg}, the analysis does not consider logarithmic systematics in the forecasting.

For more technical details about the methodology followed here see Appendix \ref{sec:Appendix-A}.

\begin{figure}[H]
  \centering
  \includegraphics[width=.49\linewidth]{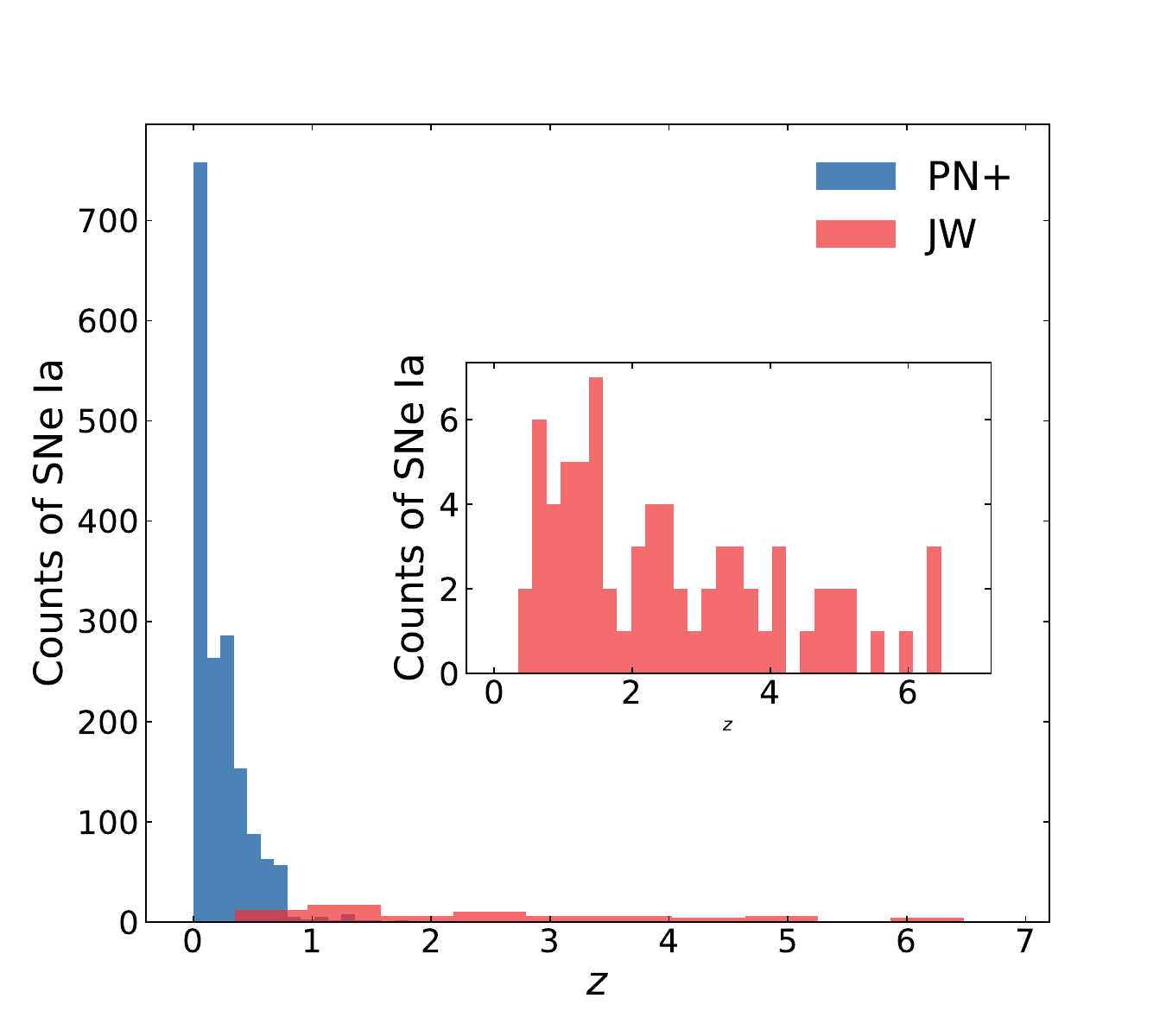}
  \includegraphics[width=.49\linewidth]{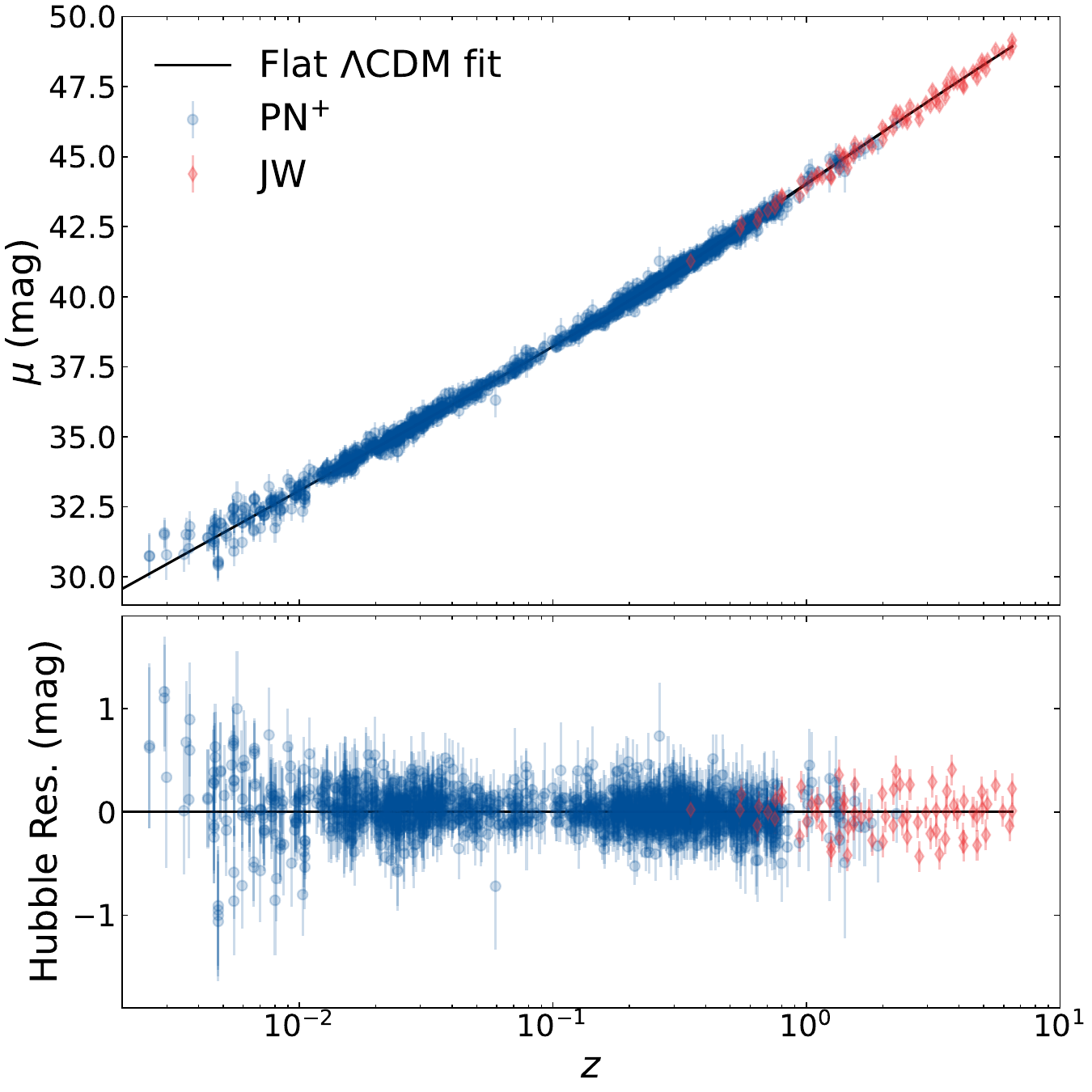}
\caption{\textit{Left:} Histogram of the Pantheon+ data (blue) and the extracted JW (red). \textit{Right:} Hubble Diagram of the Pantheon data (blue) and the extracted mock JW (red).
} 
    \label{fig:hist1}
\end{figure}

\end{itemize}


\section{Results: Cosmological constraints}
\label{sec:Results}

In this section, we discuss the constraints for the dark energy models previously described. In Tables \ref{tab:results1} and \ref{tab:results2} are reported the values for the cosmological parameters involved in each flat and non-flat model, respectively. Additionally, we used the results from the latest SH0ES measurement for the Hubble constant \cite{Riess:2021jrx} in which using the local distance ladder via Cepheid calibration was obtained $H_0 = 73.04 \pm 1.04$, and was introduced as a fixed value for the Hubble constant in our analyses. This value of $H_0$ is in accordance with which the simulated sample was constructed, and it is also based on the $\Lambda$CDM model allowing us to be consistent with the simulated sample. It is worth nothing that because the distance modulus is computed using the luminous distance (Eq.(\ref{dl(z)})), the parameter $\Gamma_0$ is degenerate with the Hubble constant $H_0$. For this reason, $H_0$ is assumed fixed and any deviation that may exist in $H_0$ falls on $\Gamma_0$. Also, an absolute magnitude $M = -19.263$ was assumed, except for flat $\Lambda$CDM where $M$ was taken as a free parameter. The optimal constraints on the cosmological parameters were derived using the \texttt{emcee} code. 
All Confidence Levels (C.L) presented in this work correspond to 68.3 and 95 $\%$ i.e., to 1 and 2$\sigma$ respectively. Finally, in the presented results the value of $\Omega_k$ is calculated directly from $\Omega_k = 1 - \Omega_m - \Omega_\Lambda$, combining the marginalized distributions of each fractional density using a \texttt{getdist}\footnote{\href{https://getdist.readthedocs.io/en/latest/}{getdist.readthedocs.io}} modified version. 


\subsection{$\Lambda$CDM model}

\begin{figure}[H]
	\centering
	\includegraphics[width=7.6cm]{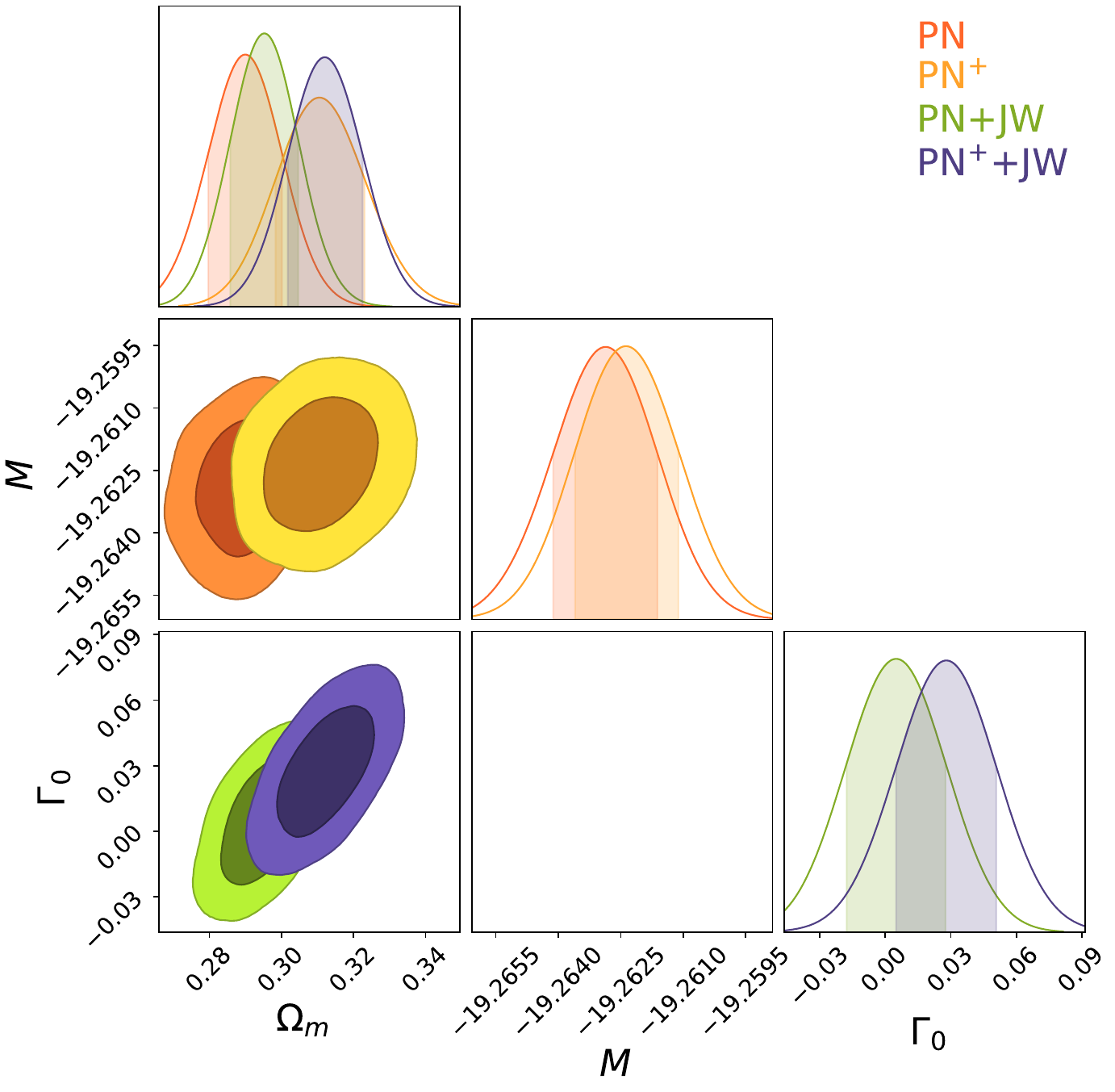}
         \includegraphics[width=7.6cm]{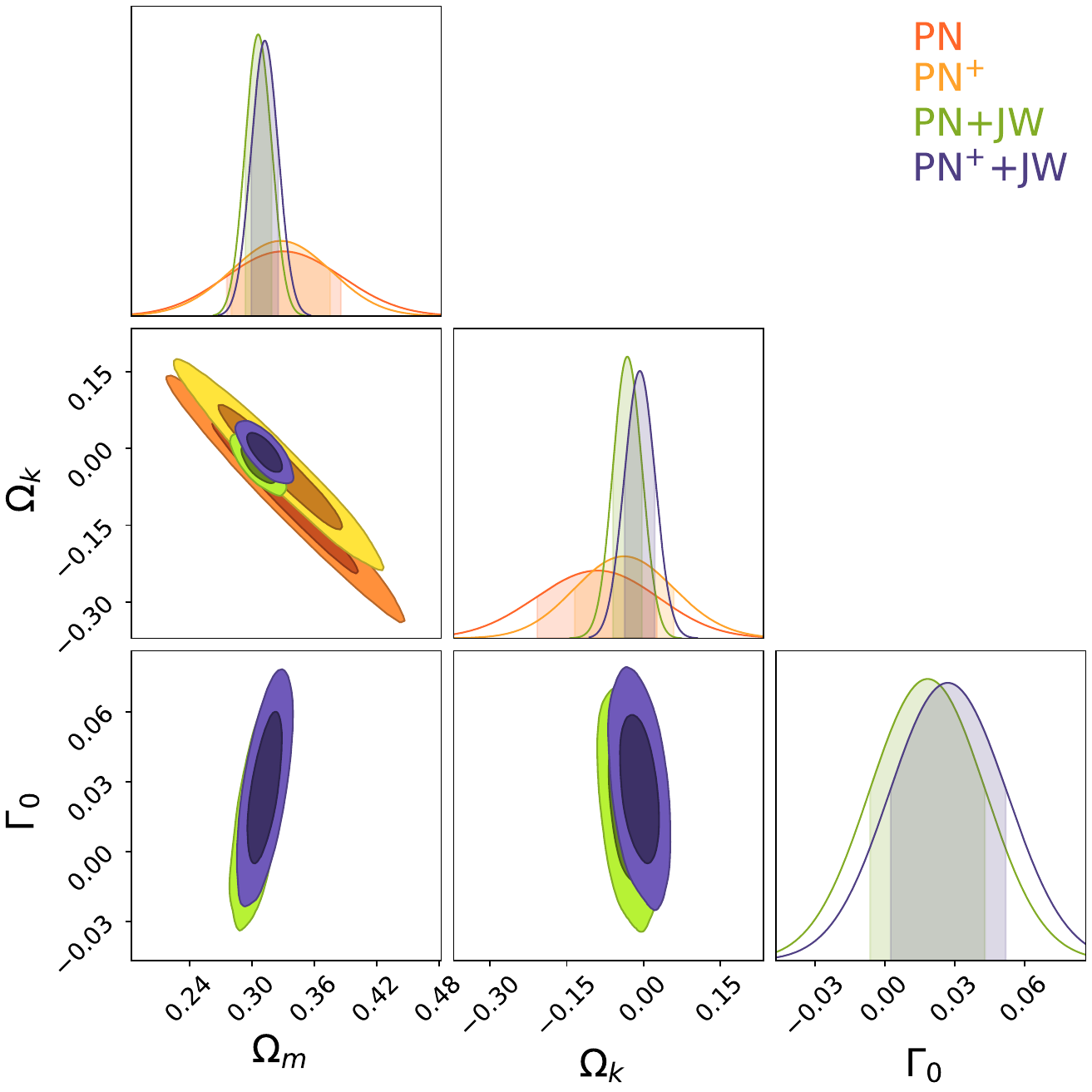}
	\caption{1-2$\sigma$ C.L results for the $\Lambda$CDM model using SNIa Pantheon--PN (red color), SNIa Pantheon+--PN${}^{+}$ (orange color), SNIa Pantheon \& JW mock data -- PN+JW (green color) and SNIa Pantheon+ \& JW mock data -- PN${}^{+}$+JW (purple color) baselines: \textit{Left:} Flat case. \textit{Right:} Non-flat case. Notice that the C.L for $M$ is not associated with $\Gamma_0$.}
	\label{fig:LCDM}
\end{figure}

The constraints for this model are given in Figure \ref{fig:LCDM}. As we can notice, using the Pantheon sample gives a relatively lower value of $\Omega_m$ than using the Pantheon+ sample for the flat-$\Lambda$CDM model. Comparing $\Omega_m = 0.290 \pm 0.008$ with $\Omega_m = 0.311^{+0.010}_{-0.009}$, shows a tendency due to the addition of JW data. Furthermore, considering the non-flat model results in a higher $\Omega_m$ estimation for the Pantheon+ sample. 

The non-flat $\Lambda$CDM model constrained by JW simulated data tends to reduce the curvature estimation towards flatness $\Omega_k \sim 0$. Let us keep in mind that the results on curvature constraints are negative for all four different dataset combinations. There is a deviation from a flat universe with $\Omega_k = -0.0092\pm 0.0091$ using the Pantheon compilation. Additionally, the flat $\Lambda$CDM model constrains $\Gamma_0$ for the JW mock sample lower than $\Gamma_0= 0.028 \pm 0.018$ mag, which implies that there is no significant difference expected for the calibrated sample.


\subsection{$w$CDM model}

\begin{figure}[H]
	\centering
	\includegraphics[width=7.6cm]{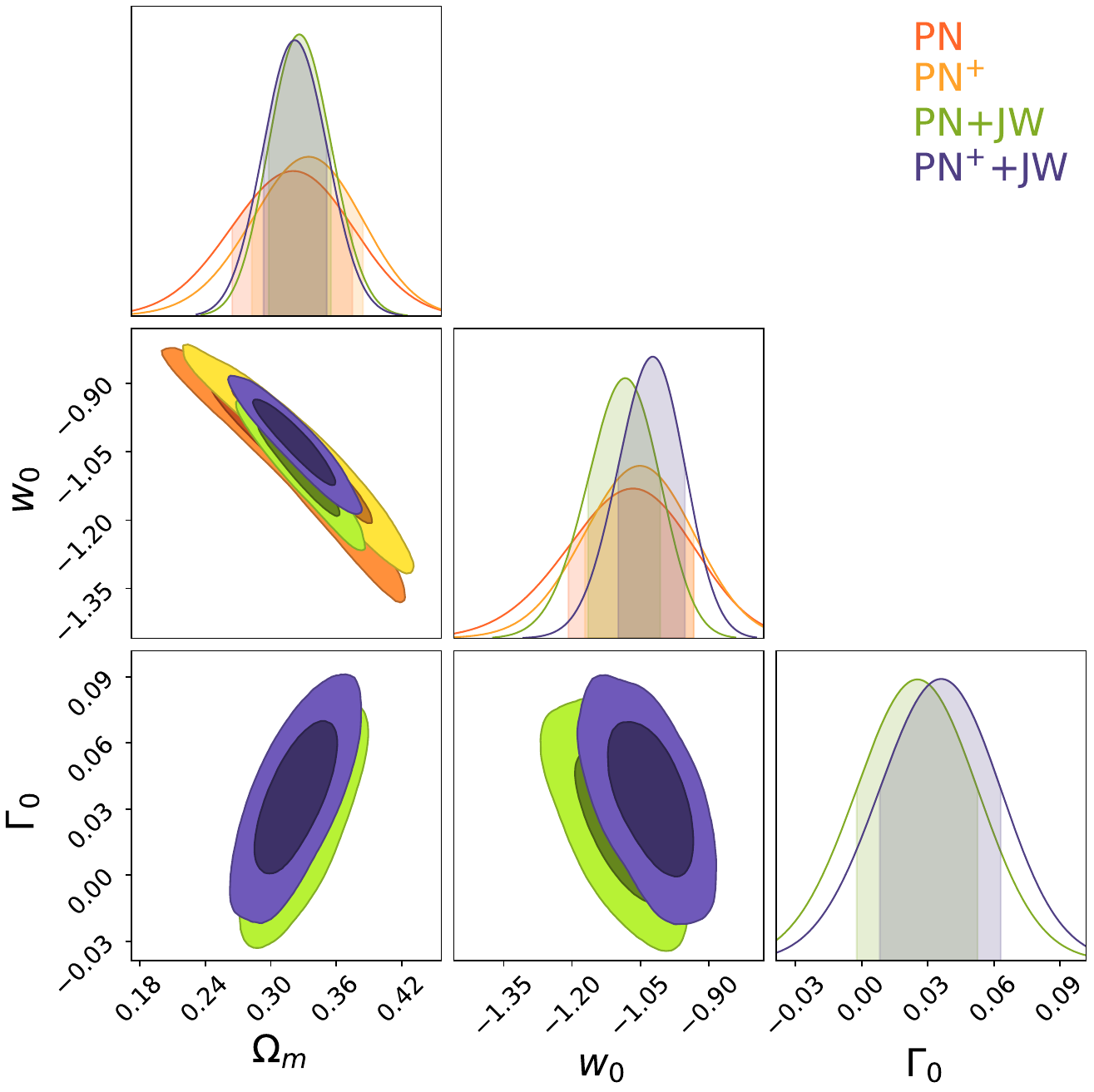}
		\includegraphics[width=7.6cm]{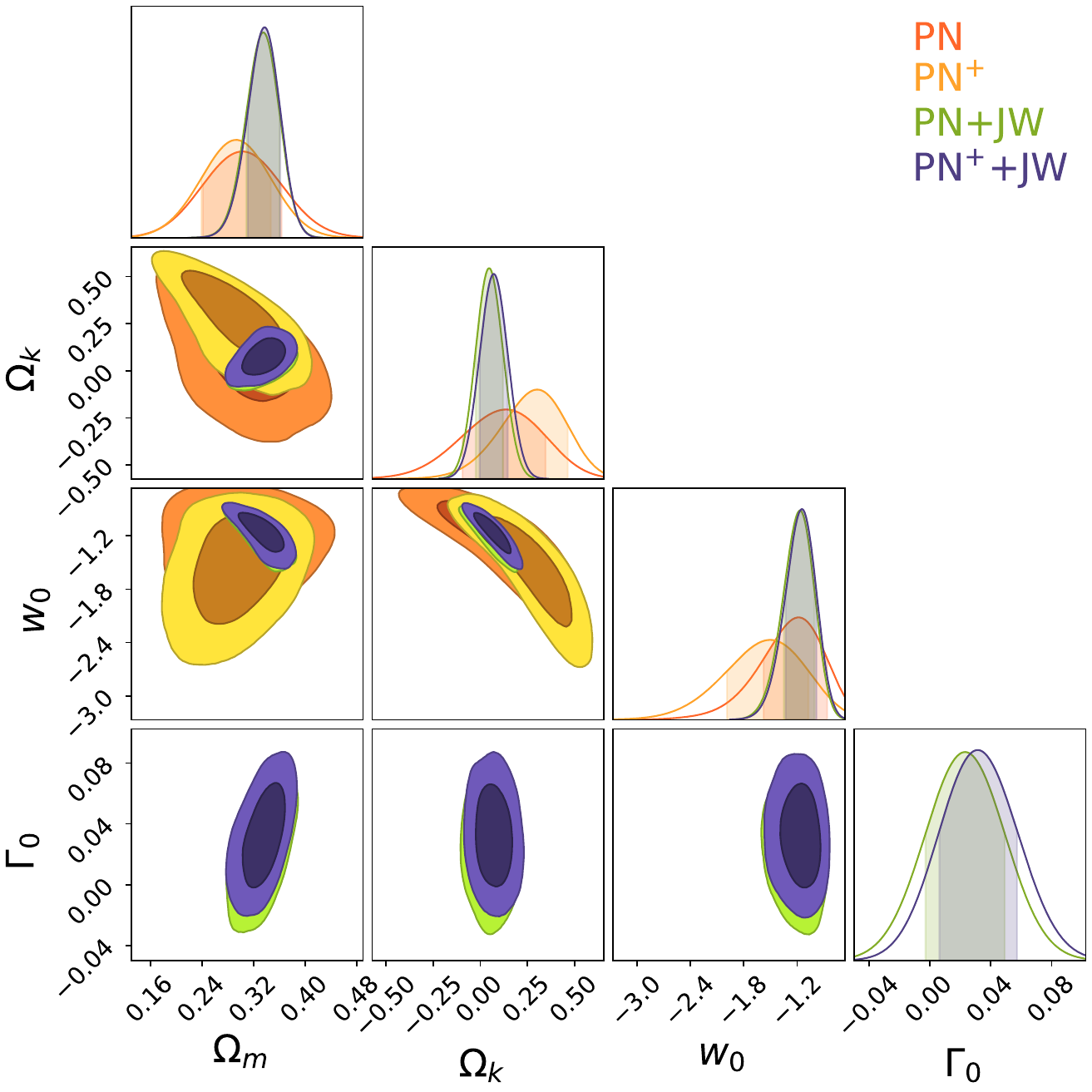}
	\caption{1-2$\sigma$ C.L results for the $w$CDM model using SNIa Pantheon--PN (orange color), SNIa Pantheon+--PN${}^{+}$ (orange color), SNIa Pantheon \& JW mock data -- PN+JW (green color) and SNIa Pantheon+ \& JW mock data -- PN${}^{+}$+JW (purple color) baselines: \textit{Left:} Flat case. \textit{Right:} Non-flat case. 
	} 
	\label{fig:wcdm}
\end{figure}

The constraints for this model are given in Figure \ref{fig:wcdm}. As we can notice, there is a correlation between the $w_0$ and $\Omega_m$ values present in the flat model, while in the non-flat version of the model, this vanishes and for the Pantheon and JW datasets there is a high error determination in the $w_0-\Omega_m$ parameter space. For this model, the fractional matter density has a similar value using Pantheon and Pantheon+ with the introduction of JW as $\Omega_m \sim 0.333$. The curvature estimation is closer to $\Omega_k =0$ as expected using the JW simulated data for the flat model. Furthermore, notice that we have a $1\sigma$ deviation from a flat model using only the Pantheon+ sample of $\Omega_k = 0.27^{+0.17}_{-0.11}$, although both SN samples prefer a non-flat universe.

Additionally, $\Gamma_0$ is constrained with a value of $\Gamma_0= 0.036^{+0.020}_{-0.022}$ mag using Pantheon+ and JW for the flat model. This means that there is not a significant deviation from the cosmological fits for the $w$CDM model using the JW mock sample compared to the observed SN samples.


\subsection{Chevallier–Polarski–Linder (CPL) model}

\begin{figure}[H]
	\centering
	\includegraphics[width=7.6cm]{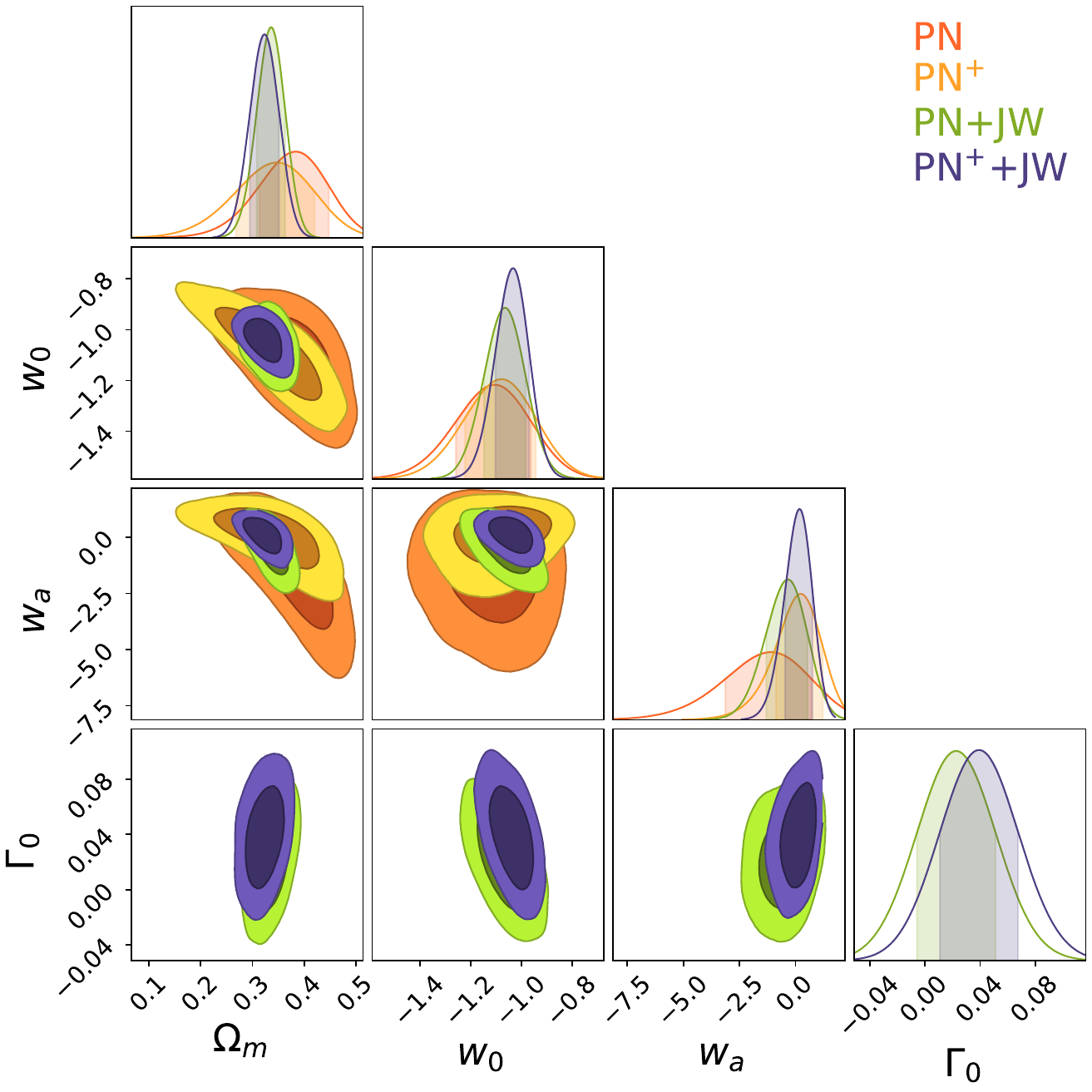}
	\includegraphics[width=7.6cm]{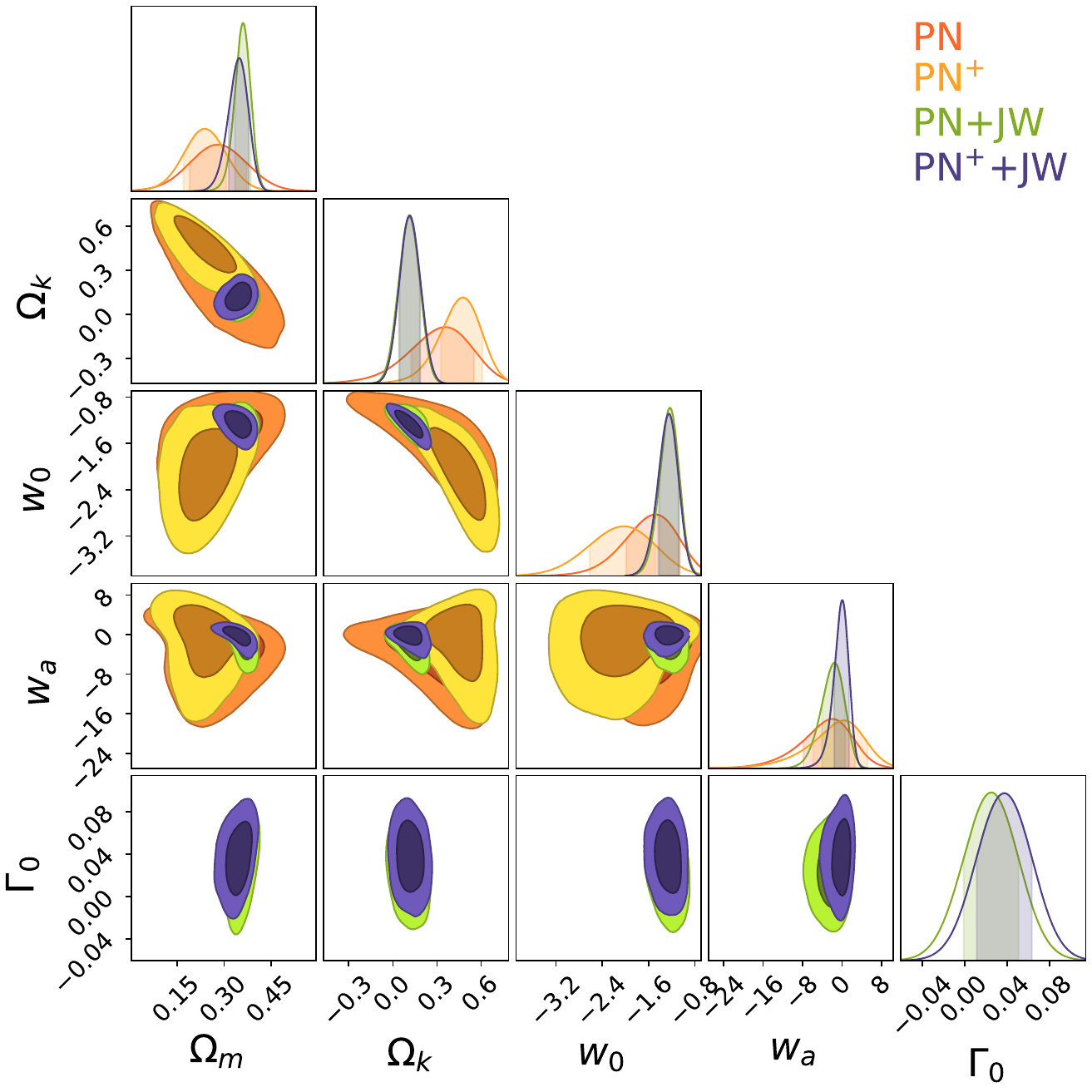}
	\caption{1-2$\sigma$ C.L results for the CPL model using SNIa Pantheon--PN (orange color), SNIa Pantheon+--PN${}^{+}$ (yellow color), SNIa Pantheon \& JW mock data -- PN+JW (green color) and SNIa Pantheon+ \& JW mock data -- PN${}^{+}$+JW (purple color) baselines: \textit{Left:} Flat case. \textit{Right:} Non-flat case. 
	}
	\label{fig:cpl}
\end{figure}

The constraints for this model are given in Figure \ref{fig:cpl}. It is interesting to notice that the errors in the cosmological parameters determined for this model using Pantheon+ are lower than the previous Pantheon SN catalog. 
This is expected due to the density of data points at lower redshifts, where the CPL model can be well-constrained. 
For the flat model and using Pantheon data, we recover the $\Lambda$CDM model with $w_0 = -1.111^{+0.110}_{-0.124}$ and $w_a = -1.252^{+1.354}_{-1.709}$ both at $1\sigma$. 
Something similar happened when we included the JW sample. This trend is confirmed when using Pantheon+ data, e.g. using Pantheon+ and JW results with $\Omega_m = 0.323^{+0.022}_{-0.023}$, $w_0 = -1.035^{+0.047}_{-0.056}$ and $w_a = 0.130^{+0.432}_{-0.37}$.
However, in the non-flat model, the estimations change when using solely SN measurements. The curvature estimation is deviated more than $1\sigma$ as for Pantheon $\Omega_k = 0.31^{+0.22}_{-0.13}$, and for Pantheon+ $\Omega_k = 0.447^{+0.14}_{-0.19}$. These results also deviate from the $\Lambda$CDM confirmation as both $w_0$ and $w_a$ do not recover the basic equations with $w_0 = -1$ and $w_a = 0$. 

In this model, all the results for $\Gamma_0$ constraints are lower than the simulated 0.15 mag error, and therefore, we do not expect to have any systematic effects on the cosmological parameters contaminated by the simulated magnitude in the JW. The larger estimation, in comparison to previous models, was found using the Pantheon+ sample for which $\Gamma_0 = 0.039 \pm 0.022$ mag.  

\subsection{Jassal–Bagla–Padmanabhan (JBP) model}

\begin{figure}[H]
	\centering
	\includegraphics[width=7.6cm]{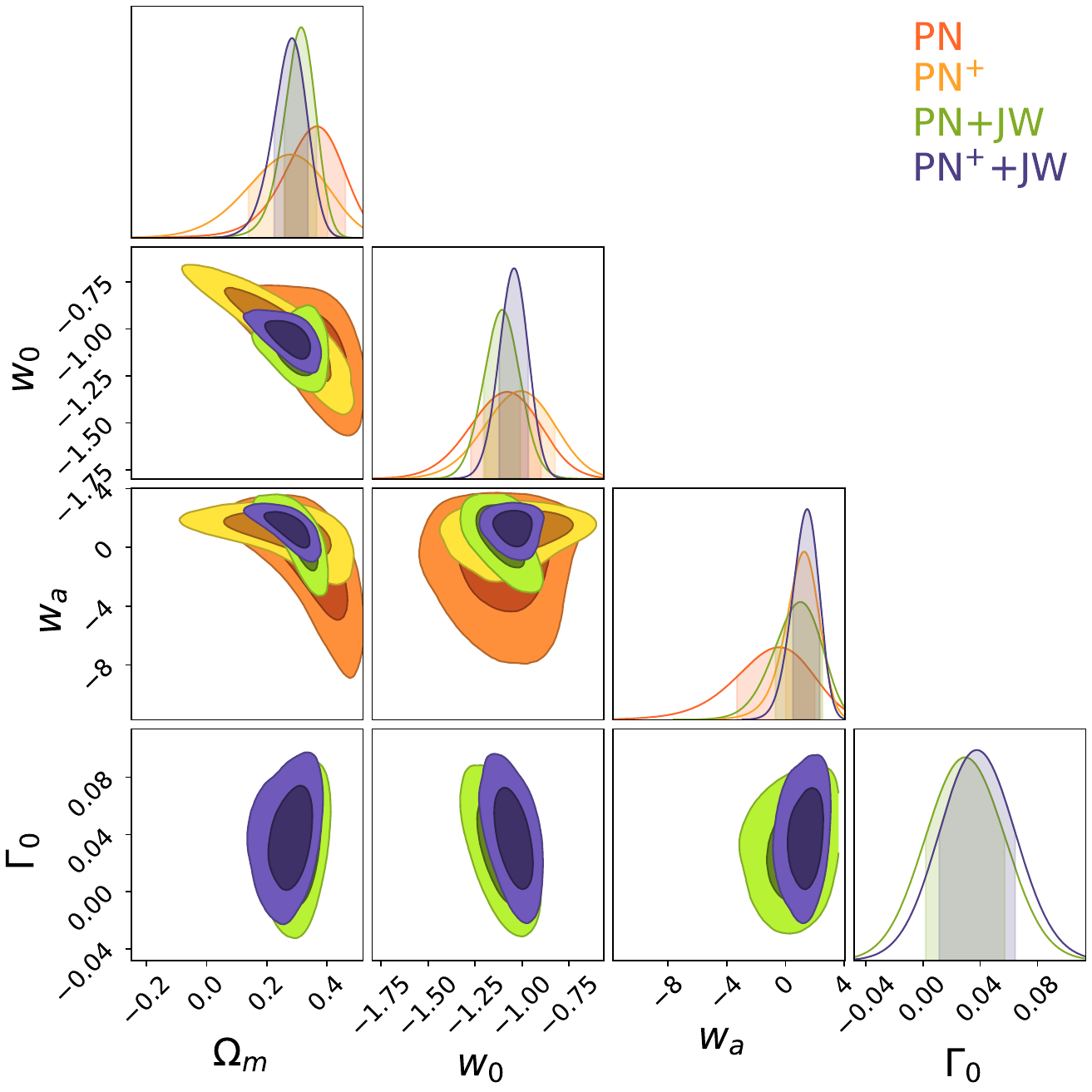}
		\includegraphics[width=7.6cm]{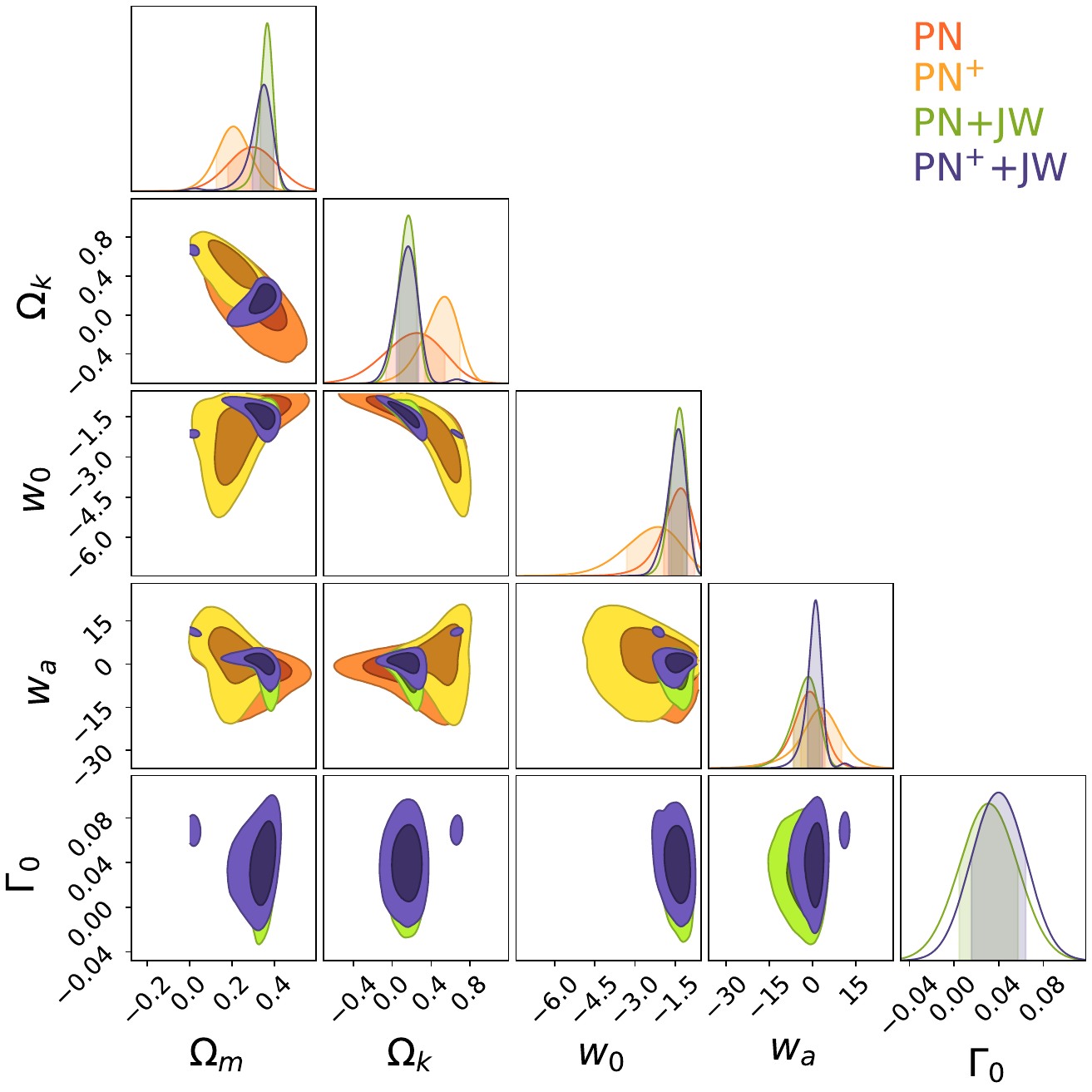}
	\caption{1-2$\sigma$ C.L results for the JBP model using SNIa Pantheon--PN (orange color), SNIa Pantheon+--PN${}^{+}$ (yellow color), SNIa Pantheon \& JW mock data -- PN+JW (green color) and SNIa Pantheon+ \& JW mock data -- PN${}^{+}$+JW (purple color) baselines: \textit{Left:} Flat case. \textit{Right:} Non-flat case. 
 }
	\label{fig:jbp}
\end{figure}

The constraints for this model are given in Figure \ref{fig:jbp}. As we can notice, the flat version of this model recovers the correlation between the $\Omega_m$ and the $w_0$ parameters for all the datasets while the $w_a$ determination is done with a large error determination.
Confirmation of $\Lambda$CDM occurs for Pantheon and JW combinations as $w_0 = -1$ and $w_a = 0$, while this not happen using Pantheon+, which results in a clear deviation as $w_a = 1.403^{+0.650}_{-0.845}$. It is worth mentioning that the only negative $w_a$ value is obtained using the Pantheon dataset as $w_a = -0.721^{+1.787}_{-2.492}$. Using Pantheon, Pantheon+ and the Pantheon and JW combination has a parameter $w_a$ determination error larger than the average. For the fractional matter density, with Pantheon data, we obtain a larger estimation than Pantheon+, with $\Omega_m > 0.3$ for Pantheon, and $\Omega_m < 0.3$ for the Pantheon+ and the combination with JW. 

For the non-flat model, we obtain different results, as none of the combinations returns a deviation from the flat model. With Pantheon+ data we obtain $\Omega_k = 0.48^{+0.20}_{-0.10}$, while the JW mock data brings the estimation closer to $\Omega_k = 0$ without reaching flatness.
It is worth noticing that Pantheon combinations result in negative $w_a$ values, while Pantheon+ combinations are opposite at $1\sigma$ with a large uncertainty. 

Using the JBP model the larger $\Gamma_0$ determination is obtained using the Pantheon+ and JW for the non-flat version as $\Gamma_0 = 0.040 \pm 0.022$ mag, which again discards the systematic error in the magnitude determination for the JW mock data.

\subsection{Exponential model}

\begin{figure}[H]
	\centering
	\includegraphics[width=7.6cm]{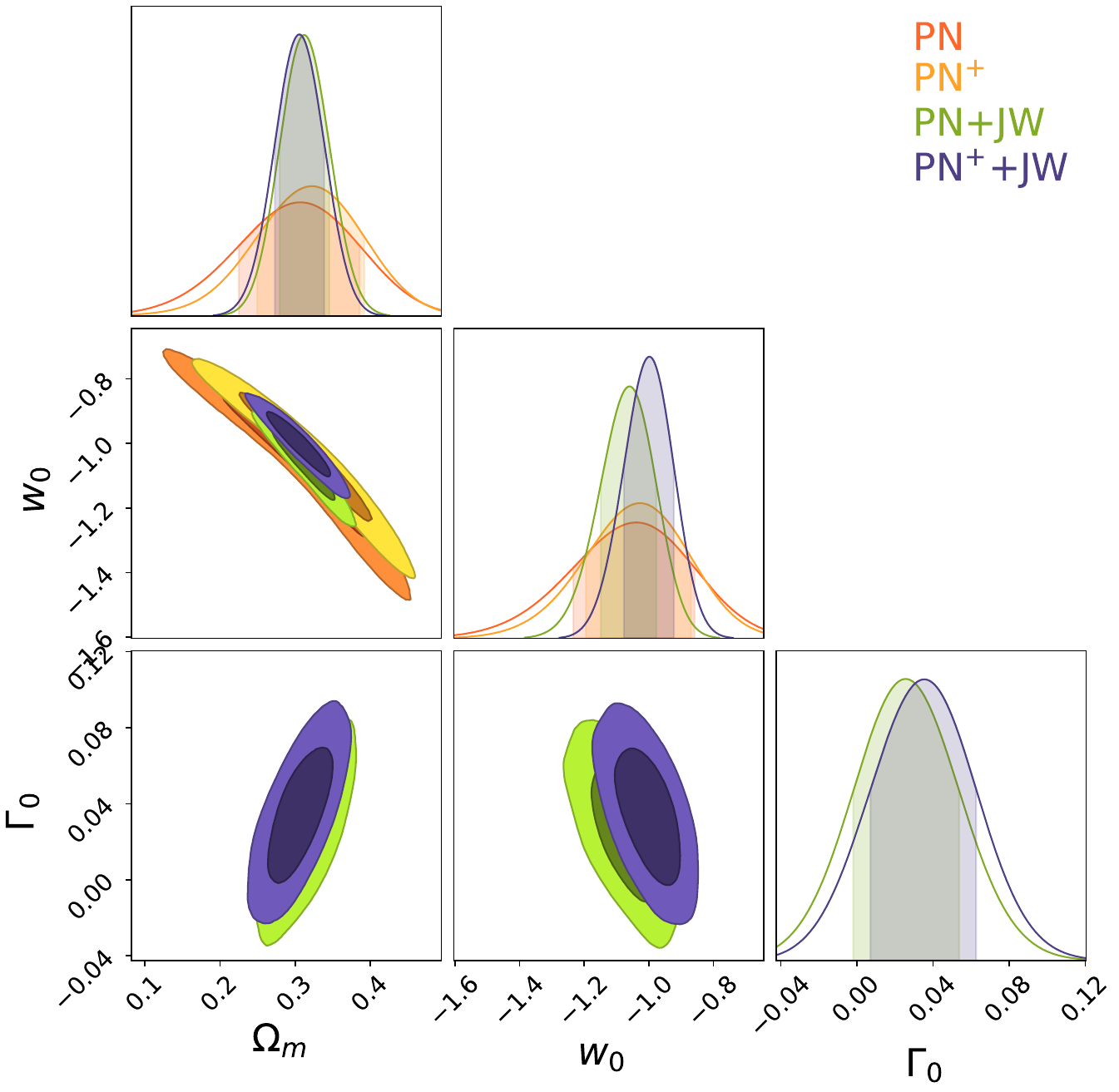}
	\includegraphics[width=7.6cm]{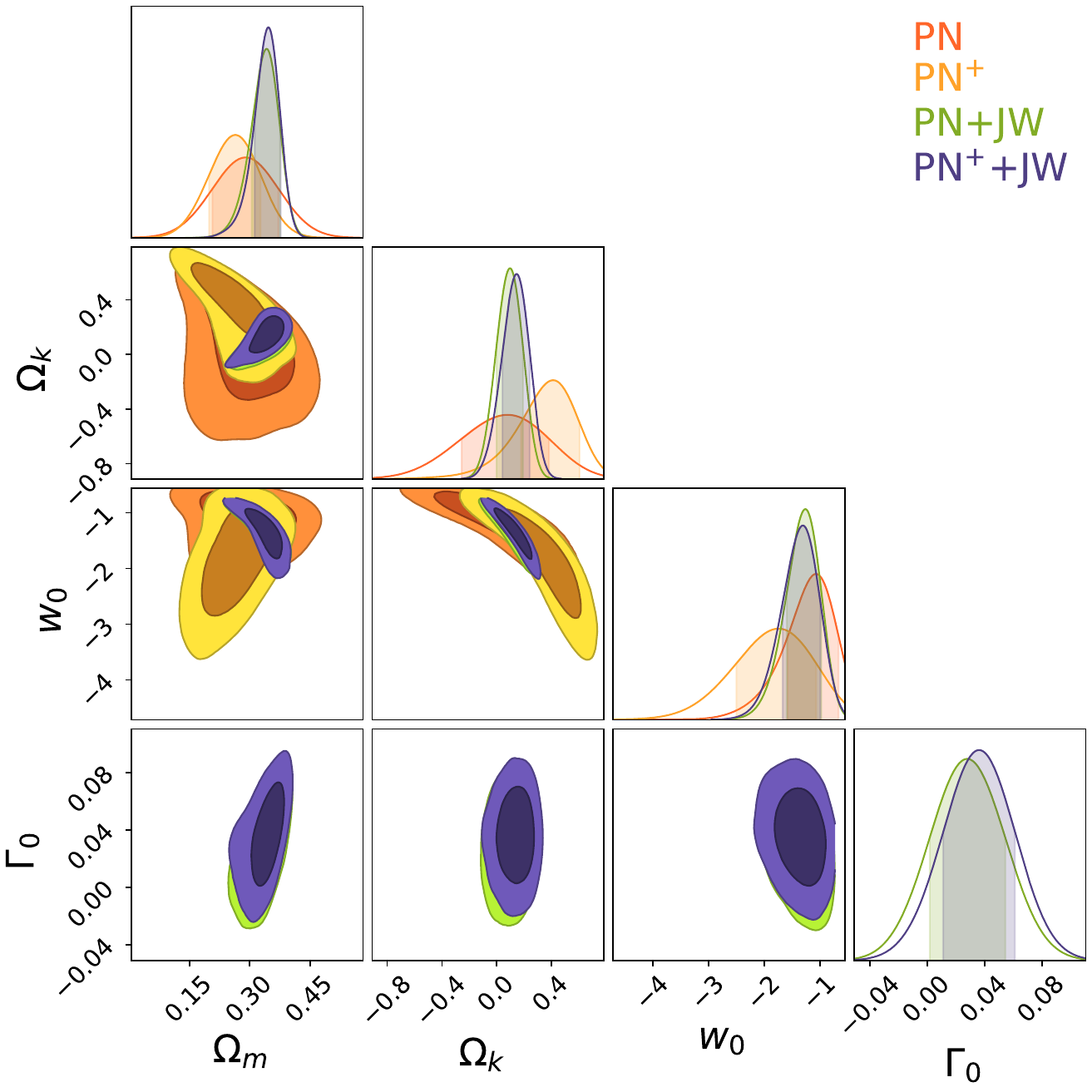}
	\caption{1-2$\sigma$ C.L results for the exponential model using SNIa Pantheon--PN (orange color), SNIa Pantheon+--PN${}^{+}$ (yellow color), SNIa Pantheon \& JW mock data -- PN+JW (blue color) and SNIa Pantheon+ \& JW mock data -- PN${}^{+}$+JW (purple color) baselines: \textit{Left:} Flat case. \textit{Right:} Non-flat case. 
 }
	\label{fig:exp}
\end{figure}

The constraints for this model are given in Figure \ref{fig:exp}. As we can notice, a correlation is presented in the parameter space between $w_0$ and $\Omega_m$ for all dataset combinations. For the flat model, all the combinations result in $w_0 \sim -1$ between $1\sigma$ with a $\Omega_m \sim 0.3$. For the non-flat version is worth noticing that all the combinations using Pantheon data result in curvature estimations close to the ones expected for a flat universe. Being closer to flatness only with Pantheon data.
Using the Pantheon+ dataset alone results in a larger deviation from the flatness as $\Omega_k = 0.35^{+0.23}_{-0.11}$.
This can be alleviated using the JW mock data as $\Omega_k = 0.135^{+0.095}_{-0.072}$ reports a larger deviation than $1\sigma$ but near to the range of the expected value for a flat universe. 
Similar to the flat model, the fractional matter density value is close to $\Omega_m \sim 0.3$, being the lowest one obtained with Pantheon+ with $\Omega_m = 0.262\pm 0.055$. In the case of $w_0$ the results are slightly lower than $w_0 = -1$, being the same Pantheon+ the lower estimation as $w_0 = -1.86^{+0.74}_{-0.48}$. 

Using the Exponential model results for the determination of $\Gamma_0 = 0.036 \pm 0.021$ in combination with Pantheon+ and JW means that the determination of the cosmological parameters in this model is not affected by the systematics in the JW mock data. 

\subsection{Barboza–Alcaniz (BA) model}

\begin{figure}[H]
	\centering
	\includegraphics[width=7.6cm]{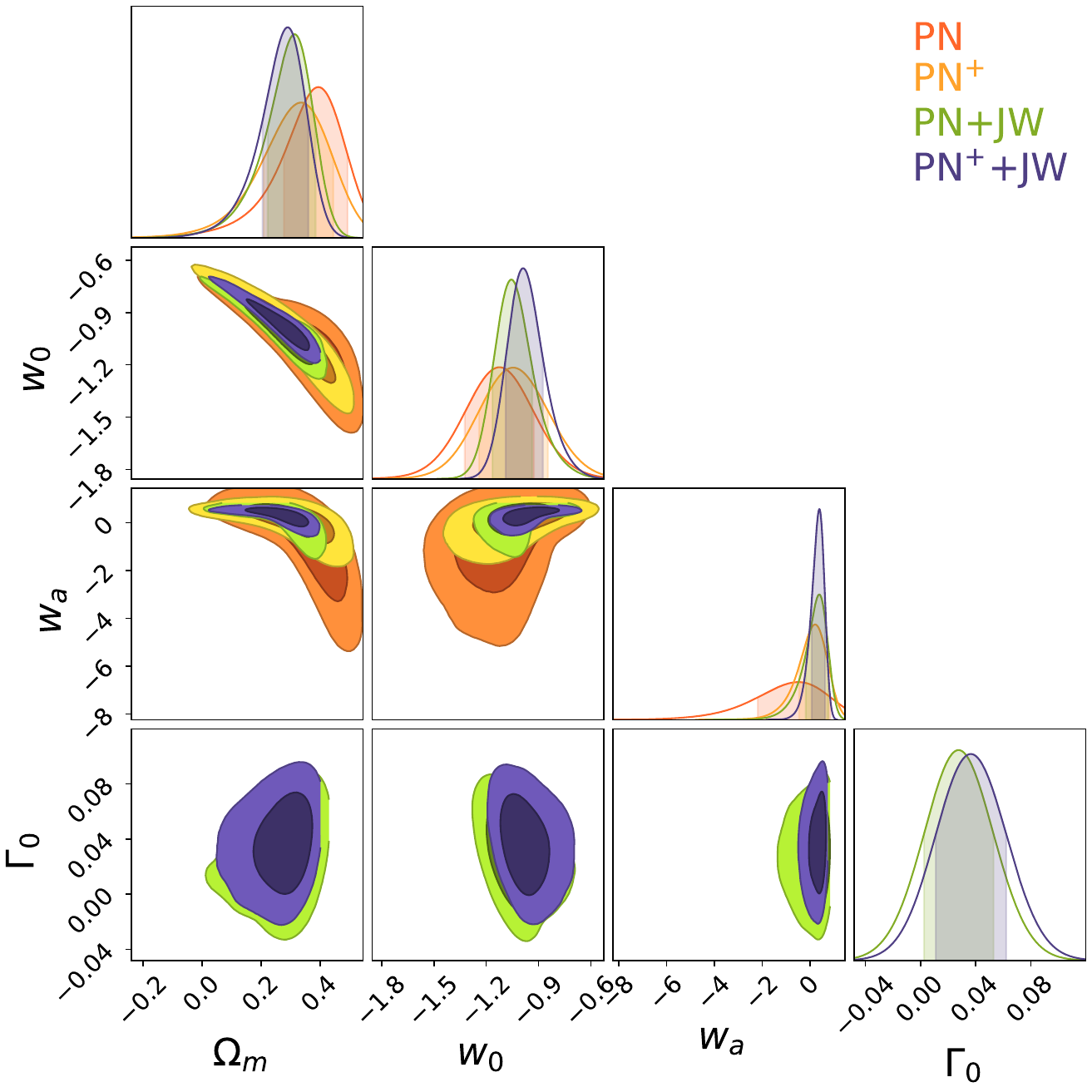}
	\includegraphics[width=7.6cm]{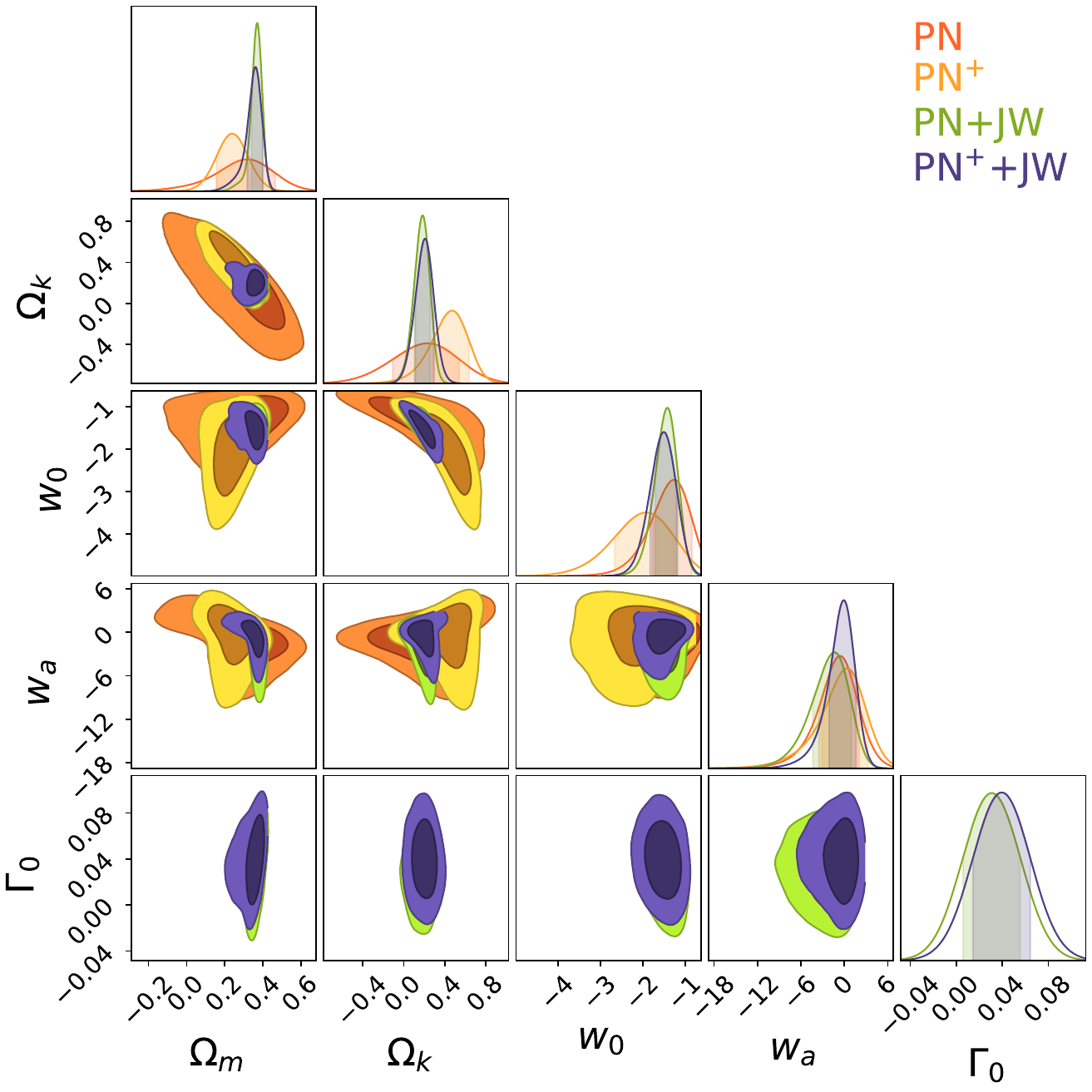}
	\caption{1-2$\sigma$ C.L results for the BA model using SNIa Pantheon--PN (orange color), SNIa Pantheon+--PN${}^{+}$ (yellow color), SNIa Pantheon \& JW mock data -- PN+JW (green color) and SNIa Pantheon+ \& JW mock data -- PN${}^{+}$+JW (purple color) baselines: \textit{Left:} Flat case. \textit{Right:} Non-flat case. 
	}
	\label{fig:ba}
\end{figure}

The constraints for this model are given in Figure \ref{fig:ba}. As we can notice for the flat model the lower value for $\Omega_m$ is the one obtained with the Pantheon+ and JW combination resulting in $\Omega_m = 0.280^{+0.047}_{-0.081}$. Regarding the values of $w_0$ and $w_a$ for the flat models, all the combination results are consistent at $1\sigma$ with the $\Lambda$CDM model, as all the results fall in the range of $w_0 \sim -1$ and $w_a =0$. Nevertheless, it is interesting to notice that the only negative value for $w_a$ is the one obtained using the Pantheon compilation with $w_a = \mathbf{-} 0.723^{+1.047}_{-1.531}$. In general, this model agrees with a flat $\Lambda$CDM at $1\sigma$. 

Meanwhile, the non-flat model shows larger deviations from the $\Lambda$CDM model as the curvature estimation is separated from the flatness in more than $1\sigma$ for all the combinations, except the Pantheon compilation as $\Omega_k = 0.20^{+0.29}_{-0.24}$. 
Other results prefer $\Omega_k > 0$.   

In this model, the larger $\Gamma_0$ estimation is obtained for the non-flat model with the Pantheon+ and JW mock data combination as $\Gamma_0 = 0.039 \pm 0.021$ mag, resulting similar to the previous models where the cosmological parameter inference is not affected by the error in magnitude for the simulated dataset. 

\section{\label{sec:conc}Conclusions}

In this paper, we studied new cosmological constraints in spatial curvature dark energy models. We extend the distance ladder method through an SNIa sample using the capabilities of JWST to forecast SNIa up to $z \sim 6$, considering the information on the star formation rates at high $z$.  
Comparing the results shown in Tables \ref{tab:results1} and \ref{tab:results2}, notice that flat $\Lambda$CDM, flat $w$CDM, and flat exponential are the only models in which the value of $\omp{m}$ obtained using the Pantheon sample is less than $\omp{m}$ for Pantheon+, i.e. $\omp{m, \text{PN}}<\omp{m, \text{PN}^{+}}$. 

However, in our analysis when including the JW mock data, the flat $\Lambda$CDM, non-flat $\Lambda$CDM, non-flat $w$CDM, and non-flat exponential are the cosmological models in which $\omp{m}$ for the combination PN+JW is less than $\omp{m}$ for PN$^{+}$+JW, i.e. $\omp{m, \text{PN+JW}}<\omp{m, \text{PN}^{+}+\text{JW}}$. Showing a lower value of this parameter when we have higher SN at $z<1$.

In regards to JW forecasting, all models have a $\Gamma_{0, \text{PN}^{+}+\text{JW}} > \Gamma_{0, \text{PN+JW}}$.
More SN at $z<1$ (e.g. Pantheon+) seems to raise the value of this $\Gamma_0$ parameter associated with JWST. Therefore, according to our $\Delta\mu$ equation (see below Eq.(\ref{eq:deltam})), it is statistically better to employ Pantheon with JW forecasting due that the vector uncertainty is lower in comparison to the one obtained with Pantheon+ catalog.
However, notice that the JW sample has been calibrated with Pantheon, therefore $\Gamma_0$ shows a preference for this SN sample. In addition, from Tables \ref{tab:results1} and \ref{tab:results2} it should be noted that the bias $\Gamma_0$ has an error less than 0.023 mag for all models. Thus, no significant changes will occur in the fit of cosmological parameters if the JW and Pantheon (Pantheon+) data sets are calibrated to within 0.023 mag.

We confirm what previous work had already shown for some cosmological models (e.g. \cite{Lu:2022utg, Salzano:2013jia}): the implementation of high redshift SN Ia data leads to the reduction of the confidence levels of the cosmological parameters. However, we have also studied the possibility of non-zero curvature in standard dark energy models. The addition of JW data improved the C.L. of the parameters $(\omp{m}, \omp{k}, w_0, w_a)$ for all models analyzed. This change being slightly greater when using Pantheon instead of Pantheon+ in almost all cases. In particular, for $\Lambda$CDM this decrease in terms of the C.L. area is a factor of $\sim 6$ for PN and $\sim 5$ for PN+.

The addition of our JW forecasting leads to an improvement in the statistics associated with $\omp{m}$ and $\omp{k}$. To give a few examples, for flat and non-flat $\Lambda$CDM model adding JW to PN+ reduces the error associated with $\omp{m}$ by $\sim 10 \%$ and $70 \%$, respectively. While the uncertainty in $\omp{k}$ is decreased by $\sim 70 \%$. As for the JBP model, the reduction in $\omp{m}$ error is $\sim 26 \%$ for the flat case and $\sim 39 \%$ for the non-flat case. Similarly, a decrease of $\sim 57 \%$ in the $\omp{k}$ error is reported.  It is expected that the JWST will observe more luminous structures with a well-treated morphology, which can help us find more robust statistics in dark energy cosmologies.


\begin{table}[H]  
\resizebox{\textwidth}{!}{%
\centering   
\begin{tabular}{lccccccccr}  
\hline\hline   
 Model  (flat)& Dataset & $\omp{M}$ & $w_0$ & $w_a$ & $\Gamma_0$ 
\\ [0.5ex]  
\hline
\\ [-2ex]
& PN & $0.290 \pm 0.008$ &  &  &  \\[0.5ex]   
& PN+JW & $0.295 \pm 0.007$ &  &  & $0.005^{+0.018}_{-0.018}$ \\[0.5ex]
\raisebox{2ex}{$\Lambda$CDM}
& PN$^{+}$ & $0.311^{+0.010}_{-0.009}$ & &  \\[0.5ex] 
& PN$^{+}$+JW & $0.312 \pm 0.008$ &  & & $0.028 \pm 0.018$ \\[2ex] 
\hline

\\ [-2ex]
& PN & $0.320^{+0.040}_{-0.045}$ & $-1.073^{+0.100}_{-0.113}$ &  &  \\[0.5ex]   
& PN+JW & $0.327^{+0.023}_{-0.021}$ & $-1.086^{+0.057}_{-0.065}$ &  & $0.025 \pm 0.021$ \\[0.5ex]
\raisebox{2ex}{ $w$CDM}
& PN$^{+}$ & $0.334^{+0.036}_{-0.041}$ &  $-1.053^{+0.091}_{-0.095}$ &  \\[0.5ex] 
& PN$^{+}$+JW & $0.322^{+0.023}_{-0.022}$ & $-1.025^{+0.052}_{-0.062}$ &  & $0.036^{+0.020}_{-0.022}$ \\[2ex] 
\hline

\\ [-2ex]
& PN & $0.380^{+0.046}_{-0.058}$ & $-1.111^{+0.110}_{-0.124}$ & $-1.252^{+1.354}_{-1.709}$ &  \\[0.5ex]   
& PN+JW & $0.336^{+0.021}_{-0.022}$ & $-1.065^{+0.061}_{-0.064}$ & $-0.399^{+0.668}_{-0.798}$ & $0.022^{+0.023}_{-0.022}$ &  \\[0.5ex]
\raisebox{2ex}{ CPL}
& PN$^{+}$ & $ 0.343^{+0.054}_{-0.065}$ & $-1.081^{+0.103}_{-0.118}$ & $0.150^{+0.638}_{-0.936}$ & \\[0.5ex] 
& PN$^{+}$+JW & $0.323^{+0.022}_{-0.023}$ & $-1.035^{+0.047}_{-0.056}$ & $0.130^{+0.432}_{-0.537}$ & $0.039 \pm 0.022$
& \\[2ex]  
\hline
\\ [-2ex]
& PN & $0.358^{+0.064}_{-0.091}$ & $-1.088^{+0.140}_{-0.162}$ & $-0.721^{+1.787}_{-2.492}$ &  \\[0.5ex]   
& PN+JW & $0.310^{+0.036}_{-0.048}$ & $-1.106^{+0.075}_{-0.079}$ & $0.873^{+1.216}_{-1.491}$ & $0.029^{+0.023}_{-0.022}$ \\[0.5ex]
\raisebox{2ex}{ JBP}
& PN$^{+}$ & $0.268^{+0.095}_{-0.121}$ & $-1.013^{+0.142}_{-0.153}$ & $1.182^{+0.707}_{-1.073}$  \\[0.5ex] 
& PN$^{+}$+JW & $0.280^{+0.039}_{-0.051}$ & $-1.045^{+0.062}_{-0.063}$ & $1.403^{+0.650}_{-0.845}$ & $0.038^{+0.022}_{-0.021}$ \\[2ex] 
\hline
\\ [-2ex]
& PN & $0.304^{+0.060}_{-0.066}$ & $-1.044^{+0.128}_{-0.163}$ &  &  \\[0.5ex]   
& PN+JW & $0.312^{+0.026}_{-0.025}$ & $-1.062^{+0.060}_{-0.073}$ &  & $0.025^{+0.022}_{-0.020}$ \\[0.5ex]
\raisebox{2ex}{ Exp}
& PN$^{+}$ & $0.321^{+0.050}_{-0.058}$ & $-1.034^{+0.118}_{-0.129}$ &  \\[0.5ex] 
& PN$^{+}$+JW & $0.306^{+0.026}_{-0.025}$ & $-1.001^{+0.058}_{-0.063}$ &  & $0.035^{+0.021}_{-0.023}$ \\[2ex] 
\hline

\\ [-2ex]
& PN & $0.383^{+0.061}_{-0.111}$ & $-1.124 \pm 0.159$ & $-0.723^{+1.047}_{-1.531}$ &  \\[0.5ex]   
& PN+JW & $0.301^{+0.049}_{-0.092}$ &  $-1.045^{+0.109}_{-0.083}$ & $0.301^{+0.254}_{-0.595}$ & $0.028 \pm 0.022$ \\[0.5ex]
\raisebox{2ex}{ BA}
& PN$^{+}$ & $0.323^{+0.076}_{-0.110}$ & $-1.044^{+0.162}_{-0.156}$ & $0.158^{+0.334}_{-0.614}$  \\[0.5ex] 
& PN$^{+}$+JW & $0.280^{+0.047}_{-0.081}$ & $-0.980^{+0.099}_{-0.075}$ & $0.343^{+0.164}_{-0.311}$ & $0.037 \pm 0.021$  \\[2ex] 
\hline

\end{tabular}  
}
\caption{Best-fit cosmological parameters at $1\sigma$ for the six flat models obtained by combining the following catalogs: Pantheon (PN) and Pantheon+ (PN$^{+}$) with the JW simulated sample (JW). Empty cells denote parameters not defined in the model.} 
\label{tab:results1}
\end{table}  


\begin{table}[H]  
\resizebox{\textwidth}{!}{%
\centering  
\begin{tabular}{lcccccccccr} 
\hline\hline   
 Model (non-flat) & Dataset & $\omp{M}$ & $\omp{\Lambda}$ & $w_0$ & $w_a$ & $\Gamma_0$ & $\omp{k}$
\\ [0.5ex]  
\hline
\\ [-2ex]
& PN & $0.332\pm 0.043$ & $0.761\pm 0.049$ &  &  &  & $-0.092\pm 0.091
$ \\[0.5ex]   
& PN+JW & $0.306\pm 0.010$ & $0.725\pm 0.017$ &  &  & $0.018\pm 0.019$ & $-0.031\pm 0.022$ \\[0.5ex]
\raisebox{2ex}{$\Lambda$CDM}
& PN$^{+}$ & $0.328\pm 0.037$ & $0.710\pm 0.041$ & & & & $-0.038\pm 0.076$  \\[0.5ex] 
& PN$^{+}$+JW & $0.312\pm 0.010$ & $0.695\pm 0.017$ &  &  & $ 0.027\pm 0.019$ & $-0.008\pm 0.023$ \\[2ex]  
\hline

\\ [-2ex]
& PN & $0.302\pm 0.051$ & $0.59^{+0.11}_{-0.20}$ & $-1.29^{+0.38}_{-0.18}$ &  &  & $0.11^{+0.21}_{-0.16}$ \\[0.5ex]   
& PN+JW & $0.333^{+0.022}_{-0.019}$ & $ 0.619^{+0.066}_{-0.073}$ & $-1.19^{+0.16}_{-0.11}$ &  & $0.023\pm 0.021$ & $0.048\pm 0.058$ \\[0.5ex]
\raisebox{2ex}{$w$CDM}
& PN$^{+}$ & $0.292\pm 0.045$ & $0.435^{+0.068}_{-0.15}$ & $-1.58^{+0.46}_{-0.29}$ & & & $0.27^{+0.17}_{-0.11}$  \\[0.5ex] 
& PN$^{+}$+JW & $0.334^{+0.022}_{-0.018}$ & $0.595^{+0.065}_{-0.079}
$ & $-1.158^{+0.124}_{-0.152}$ &  & $0.032\pm 0.021$ & $0.071\pm 0.061$ \\[2ex]  
\hline

\\ [-2ex]
& PN & $0.280 \pm 0.077$ & $0.412^{+0.060}_{-0.16}$ & $-1.537^{+0.346}_{-0.451}$ & $-3.6^{+5.4}_{-2.9}$ & & $0.31^{+0.22}_{-0.13}$ \\[0.5ex]   
& PN+JW & $0.358^{+0.022}_{-0.017}$ & $0.533^{+0.054}_{-0.077}$ & $-1.238^{+0.126}_{-0.143}$ & $-2.0^{+2.4}_{-1.3}$ & $0.025 \pm 0.021$ & $0.109^{+0.061}_{-0.055}$ \\[0.5ex]
\raisebox{2ex}{ CPL}
& PN$^{+}$ & $0.239\pm 0.059$ & $0.314^{+0.038}_{-0.11}$ & $-2.059^{+0.465}_{-0.520}$ & $-1.6^{+5.7}_{-2.7}$ & & $0.447^{+0.14}_{-0.09}$ \\[0.5ex] 
& PN$^{+}$+JW & $0.344^{+0.029}_{-0.020}$ & $0.539^{+0.069}_{-0.070}$ & $-1.17^{+0.16}_{-0.11}$ & $-0.34^{+1.5}_{-0.71}$ & $0.037 \pm 0.021$ & $0.116\pm 0.057$ \\[2ex] 
\hline

\\ [-2ex]
& PN & $0.292\pm 0.097$ & $ 0.50^{+0.12}_{-0.24}$ & $-1.50^{+0.62}_{-0.21}$ & $-2.1^{+4.8}_{-2.6}$ & & $0.21^{+0.29}_{-0.21}$ \\[0.5ex]   
& PN+JW & $0.358^{+0.031}_{-0.016}$ & $0.493^{+0.055}_{-0.12}$ & $-1.372^{+0.225}_{-0.281}$ & $-3.0^{+4.9}_{-2.1}$ & $0.031 \pm 0.022$ & $ 0.149^{+0.092}_{-0.061}$ \\[0.5ex]
\raisebox{2ex}{JBP}
& PN$^{+}$ & $0.204\pm 0.072$ & $0.315^{+0.044}_{-0.15}$ & $-2.320^{+0.781}_{-1.051}$ & $2.1^{+6.7}_{-4.3}$ & & $0.48^{+0.20}_{-0.10}$  \\[0.5ex] 
& PN$^{+}$+JW & $0.328^{+0.056}_{-0.012}$ & $0.517^{+0.076}_{-0.16}$ & $-1.424^{+0.277}_{-0.347}$ & $0.5^{+2.4}_{-1.4}$ & $0.040 \pm 0.022$ & $0.155^{+0.10}_{-0.088}$ \\[2ex] 
\hline

\\ [-2ex]
& PN & $0.289\pm 0.068$ & $0.67^{+0.16}_{-0.32}$ & $-1.24^{+0.50}_{-0.18}$ &  &  & $0.04^{+0.30}_{-0.24}$ \\[0.5ex]   
& PN+JW & $0.335^{+0.035}_{-0.021}$ & $0.573^{+0.08}_{-0.12}$ & $-1.33^{+0.31}_{-0.19}$ &  & $0.028\pm 0.022$ & $0.092^{+0.086}_{-0.076}$ \\[0.5ex]
\raisebox{2ex}{ Exp}
& PN$^{+}$ & $ 0.262\pm 0.055$ & $0.391^{+0.046}_{-0.18}$ & $-1.86^{+0.74}_{-0.48}$ & & & $0.35^{+0.23}_{-0.11}$  \\[0.5ex] 
& PN$^{+}$+JW & $0.340^{+0.032}_{-0.018}$ & $0.526^{+0.073}_{-0.12}$ & $-1.38^{+0.34}_{-0.22}$ &  & $0.036 \pm 0.021$ & $0.134^{+0.095}_{-0.072}$ \\[2ex] 
\hline

\\ [-2ex]
& PN & $0.287^{+0.15}_{-0.086}$ & $0.52^{+0.11}_{-0.23}$ & $-1.328^{+0.324}_{-0.459}$ & $-1.1^{+2.6}_{-1.7}$ &  & $0.20^{+0.29}_{-0.24}$ \\[0.5ex]   
& PN+JW & $0.363^{+0.029}_{-0.014}$ & $0.465^{+0.044}_{-0.090}$ & $-1.447^{+0.206}_{-0.255}$ & $-2.2^{+2.9}_{-1.3}$ & $0.031 \pm 0.022$ & $0.172^{+0.075}_{-0.054}$ \\[0.5ex]
\raisebox{2ex}{ BA}
& PN$^{+}$ & $0.240 \pm 0.069$ & $0.332^{+0.046}_{-0.13}$ & $-1.990^{+0.556}_{-0.658}$ & $-0.7^{+3.2}_{-1.3}$ & & $0.43^{+0.19}_{-0.10}$ \\[0.5ex] 
& PN$^{+}$+JW & $0.347^{+0.042}_{-0.012}$ & $0.453^{+0.053}_{-0.10}$ & $-1.522^{+0.254}_{-0.280}$ & $-0.69^{+1.8}_{-0.92}$ & $0.039 \pm 0.021$ & $0.200^{+0.081}_{-0.070}$ \\[2ex] 
 
\hline
\end{tabular}  
}
\caption{Best-fit cosmological parameters at $1\sigma$ for the six non-flat cosmological models obtained for the following catalogs: combining Pantheon (PN) and Pantheon+ (PN$^{+}$) datasets with the JW simulated sample (JW).  Empty cells denote parameters not defined in the model..}  
\label{tab:results2}
\end{table}


\begin{acknowledgments}
The Authors thank J. Vinko and E. Regős for their insights on using JWST forecasting data. Also, we would like to acknowledge funding from PAPIIT UNAM Project TA100122. 
CE-R acknowledges the Royal Astronomical Society as FRAS 10147.
PMA and RS are supported by the CONACyT National Grant.
The computational calculations have been carried out using facilities procured through the Cosmostatistics National Group project.
This article is based upon work from COST Action CA21136 Addressing observational tensions in cosmology with systematics and fundamental physics (CosmoVerse) supported by COST (European Cooperation in Science and Technology). 
\end{acknowledgments}


\appendix

\section{SNIa JWST baseline forecasting: data and priors} 
\label{sec:Appendix-A}

The simulated data set used is derived from the FLARE project, which has the goal of searching supernovas from population III at redshift $z \geq 10$ \cite{Wang:2017awy} by using the characteristics of the JWST in an area of 0.05 square degrees. Furthermore, the project employs four broadband in the NIRCAM filters (F150W, F200W, F322W2, F444W) with exposure times that can reach 10$\sigma$ limiting magnitudes of $m \gtrsim 27$ in these filters \cite{Lu:2022utg}. By using Monte Carlo methods it was found that for a specific project of JWST observation, at least 200 SNe Ia could be observed. Therefore, the mock sample is constructed using a flat $\Lambda$CDM model with $H_0 = 71.66$ km s$^{-1}$ Mpc$^{-1}$ and $\Omega_m = 0.3$, as derived using Pantheon data \cite{Scolnic:2019apa}. To ensure consistency between the local and distant samples, we consider a Gaussian error associated with the supernovae distance modules of 0.15 mag 
and extrapolated them to a larger uncertainty with higher redshift, although this is an oversimplification of the typical distance measurements \cite{Phillips1993ApJ}. 
For the FLARE project \cite{Regos:2018fgq} the simulations were done using the standard cosmology and a different SFR according to the redshift in which the simulation is done. Also, it is considered the calculation of the rate of occurrence to estimate the range of the detection needed in the telescope to perform the observations. 

The used SFR functions are explicitly proposed to have a redshift dependence. For low $z \lesssim 3$, the function used is \cite{Hopkins_2006}:  
\begin{equation}
    \mathrm{SFR}(z) = K \frac{(a+bz)h}{1+ (z/c)^d},
\end{equation}
where $h = H_0/(100 \mathrm{km} \mathrm{s}^{-1} \mathrm{Mpc}^{-1})$, $a= 0.017$, $b = 0.13$, $c = 3.3$, and $d = 5.3$. These quantities constrain $K$ with the SN rates observed. For higher redshifts between $3 < z \leq 8$, the function used is: 
\begin{equation}
    \mathrm{SFR}(z) \propto (1+z)^{-3.6},
\end{equation}
and for $z > 8$:
\begin{equation}
    \mathrm{SFR}(z) \propto (1+z)^{-10.4}.
\end{equation}
Additionally, other SFRs are proposed for the whole interval \cite{Madau:2014bja}: 
\begin{equation}
    \mathrm{SFR}(z) = 0.015 \frac{(1+z)^{2.7}}{1 + [(1+z)/2.9]^{5.6}},
\end{equation}
or \cite{behroozi2013average}:
\begin{equation}
    \mathrm{SFR}(z) = \frac{C}{10^{A(z-z_0)} + 10^{B(z-z_0)}},
\end{equation}
using the constants $C = 0.18$, $A = -0.997$, $B = 0.241$, and $z_0 = 1.243$. All the parameterisations studied in Sec.\ref{sec:darkenergy} for the SFR give similar redshift dependence with a peak in star formation in $z \sim 2$. 
The volume rate was calculated using the observed SN rate per redshift bin as \cite{Regos:2018fgq}: 
\begin{equation}
    \dot{N}_{\mathrm{SN}}(z) = \frac{\dot{n}_{\mathrm{SLSN}}(z)}{1+z} \dv{V}{z},
\end{equation}
where the comoving volume can be rewritten as: 
\begin{equation}
    \dot{n}_{\mathrm{SLSN}}(z) = \varepsilon(z) \mathrm{SFR}(z),
\end{equation}
where explicitly is expressed the comoving rate of SNe. $\varepsilon(z)$ is a factor of the efficiency taking into account the metallicity dependence. Even though, $\varepsilon(z)$ can be studied through GRBs \cite{Regos:2018fgq}. So, the expected number of SN can be calculated as: 
\begin{equation}
    N_{\mathrm{SN}} = \Omega T \int_{z}^{z + \Delta z} \frac{\dot{n}_{\mathrm{SLSN}}(z)}{1+z} \dv{V}{z} \mathrm{d}z,
\end{equation}
resulting in a number per unit of redshift interval in the survey area. $\Omega$ is the survey area and $T$ is the dedicated time of observation. The redshift dependence also has to consider the number of SNIa progenitors that occur (as SNIa needs a White Dwarf) and the delay for the stellar evolution in such binary systems. 

For the simulations of mock data sets the code developed by \cite{barbary-software} is used to create light curves at different redshifts taking into account the calculated rate and detection of the observatory \cite{Regos:2018fgq}. So, for the simulation the redshift $z$, the luminosity distance $D_L$, maximum light moment $t_{\mathrm{max}}$, V absolute magnitude $M_\mathrm{V}$, stretch and colour are calculated for every SN. It is important to mention that the assumptions of the FLARE project imply that the JSWT will take deep observations for at least three years with a 90-day cadence allowing us to discover between 5 and 20 supernovae events meaning at least fifty in redshift $1 < z < 4$ \cite{Lu:2022utg}. Additionally, the simulation takes into account that the SNIa ideal observation time has to be from 2 weeks before the maximum up to one month after in which spectrum would have the ideal quality \cite{Hu:2022ddz}. So, for the simulations, the results are the Hubble diagram for the apparent magnitude assuming detection with the NIRCam of the telescope.

\bibliographystyle{JHEP}
\bibliography{references}

\end{document}